\documentclass[10pt,journal,compsoc]{IEEEtran}
\ifCLASSOPTIONcompsoc
  \usepackage[nocompress]{cite}
\else
  \usepackage{cite}
\fi
\ifCLASSINFOpdf
\else
\fi
\usepackage{graphicx}
\usepackage[ruled,vlined]{algorithm2e}
\usepackage{amsfonts}
\usepackage{booktabs}                  
\usepackage{gensymb}
\usepackage{amsmath}
\usepackage{comment}
\usepackage{caption}
\usepackage{bm}
\usepackage{cclicenses}
\usepackage{makecell}
\usepackage{lscape,array}
\usepackage{multirow}
\usepackage{color} 
\usepackage{anyfontsize}
\graphicspath{{./}{./figures/}}
\usepackage{tikz}

\usepackage[bookmarks,backref=true,linkcolor=black]{hyperref} 
\hypersetup{
  pdfauthor = {},
  pdftitle = {},
  pdfsubject = {},
  pdfkeywords = {},
  colorlinks=true,
  linkcolor= black,
  citecolor= black,
  pageanchor=true,
  urlcolor = black,
  plainpages = false,
  linktocpage
}

\makeatletter
\newcommand\notsotiny{\@setfontsize\notsotiny\@vipt\@viipt}
\makeatother
\usepackage{anyfontsize}
\usepackage{t1enc}

\usepackage{color} 
\usepackage{colortbl}
\usepackage{cases}
\definecolor{light-gray}{gray}{0.6}
\definecolor{lavender}{rgb}{0.5,0.5,1.0}

\usepackage{verbatim}
\usepackage{url}
\usepackage{gensymb}

\definecolor{carmine}{rgb}{0.59, 0.0, 0.09}
\definecolor{richcarmine}{rgb}{0.84, 0.0, 0.25}
\definecolor{mediumcarmine}{rgb}{0.69, 0.25, 0.21}
\definecolor{carminered}{rgb}{1.0, 0.0, 0.22}
\definecolor{ruddypink}{rgb}{0.88, 0.56, 0.59}
\definecolor{brinkpink}{rgb}{0.98, 0.38, 0.5}
\definecolor{cherryblossompink}{rgb}{1.0, 0.72, 0.77}
\definecolor{electricblue}{rgb}{0.49, 0.98, 1.0}
\definecolor{gold(web)(golden)}{rgb}{1.0, 0.84, 0.0}
\definecolor{green(pigment)}{rgb}{0.0, 0.65, 0.31}
\definecolor{green-yellow}{rgb}{0.68, 1.0, 0.18}
\definecolor{azure(colorwheel)}{rgb}{0.0, 0.5, 1.0}
\definecolor{electricpurple}{rgb}{0.75, 0.0, 1.0}
\newcommand{\tikzcircle}[2][black,fill=red]{\tikz[baseline=-0.5ex]\draw[#1,radius=#2] (0,0) circle ;}

\SetKwInput{KwInput}{Input}                
\SetKwInput{KwOutput}{Output}              

\begin{document}
%
\title{DL4SciVis: A State-of-the-Art Survey on\\ Deep Learning for Scientific Visualization}
%
%

\author{Chaoli~Wang,~\IEEEmembership{Senior~Member,~IEEE} and Jun Han
\IEEEcompsocitemizethanks{\IEEEcompsocthanksitem C. Wang and J. Han are with the Department of Computer Science and Engineering, University of Notre Dame, Notre Dame, IN 46556. E-mail: \{chaoli.wang, jhan5\}@nd.edu. \protect
}
}


%
%
%
\markboth{IEEE Transactions on Visualization and Computer Graphics}{Wang \MakeLowercase{\textit{et al.}}: DL4SciVis: A State-of-the-Art Survey on Deep Learning for Scientific Visualization}
%



\IEEEtitleabstractindextext{%
\begin{abstract}
Since 2016, we have witnessed the tremendous growth of artificial intelligence+visualization (AI+VIS) research. However, existing survey papers on AI+VIS focus on visual analytics and information visualization, not scientific visualization (SciVis). In this paper, we survey related deep learning (DL) works in SciVis, specifically in the direction of DL4SciVis: designing DL solutions for solving SciVis problems. To stay focused, we primarily consider works that handle scalar and vector field data but exclude mesh data. We classify and discuss these works along six dimensions: domain setting, research task, learning type, network architecture, loss function, and evaluation metric. The paper concludes with a discussion of the remaining gaps to fill along the discussed dimensions and the grand challenges we need to tackle as a community. This state-of-the-art survey guides SciVis researchers in gaining an overview of this emerging topic and points out future directions to grow this research. 
\end{abstract}


}

\maketitle

\IEEEdisplaynontitleabstractindextext

%

\IEEEraisesectionheading{\section{Introduction}\label{sec:introduction}}

\IEEEPARstart{B}{ack} in 2007, Ma~\cite{Ma-CGA07} pointed out the great potential of machine learning (ML) techniques to boost the next generation of visualization (VIS) research. However, before 2010, only sporadic research efforts applied artificial neural networks (ANN) to solve visualization problems, such as volume classification~\cite{Tzeng-VIS03,Tzeng-TVCG05} and flow field modeling~\cite{Kimura-JOV09}. 		
As part of ML techniques, deep learning (DL) uses multiple layers in the ANN design to extract higher-level features from data. 
With the impressive advances in graphics hardware~\cite{Raina-ICML09,Ciresan-NC10,Sze-IEEE17} and DL architectures (e.g., AlexNet~\cite{Krizhevsky-NIPS12} for image classification, GAN~\cite{Goodfellow-NIPS14} for image generation, and U-Net~\cite{Ronneberger-MICCAI15} for image segmentation), 
the renaissance of artificial intelligence (AI) as a viable solution for solving challenging problems has quickly swept across a wide variety of fields. 

Emerged in 2016, AI+VIS has quickly become the fastest-growing area in VIS, with a foreseeable impact for decades to come. Generally speaking, there are two AI+VIS directions: AI4VIS (i.e., designing AI solutions for solving VIS problems) and VIS4AI (i.e., applying VIS techniques for explainable AI). 
We refer interested readers to recent surveys on AI+VIS~\cite{Endert-CGF17,Liu-VI17,Hohman-TVCG19,Chatzimparmpas-CGF20,Yuan-CVM21,Wu-TVCG,Wang-TVCG,Alicioglu-CG22} to gain a comprehensive overview of this research area. 
These prior surveys focus on visual analytics (VA) and information visualization (InfoVis). 
A recent survey of visualization in astrophysics~\cite{Lan-CGF21} briefly discusses the use of ML techniques for scientific visualization (SciVis) but is restricted to the particular application of focus. 
In contrast, this survey studies recent advances in DL for SciVis. 

DL techniques can bring crucial benefits to SciVis. For example, inference can be performed more efficiently than conventional methods once a neural network is trained~\cite{Gu-CGA21}. Furthermore, DL solutions offer a performance boost, such as data interpolation quality~\cite{Han-VIS21}, reduction rate~\cite{Lu-CGF21}, or segmentation accuracy~\cite{Wang-VIS20}, compared with non-DL solutions. 
SciVis data and tasks share significant similarities with those in computer vision (CV) and computer graphics (CG). 
Typical SciVis data are 3D and time-dependent scalar and vector volumes, resembling their 2D image and video counterparts in CV and 3D models and animations in CG. 
CV and CG tasks (e.g., feature learning, extraction, and tracking; data classification, segmentation, generation, and prediction) can easily find their place in SciVis. 
Visualizing 3D volumetric data brings the same lighting and viewpoint optimization issues as rendering 3D models in CG. 
Therefore, it is not surprising that the development of DL solutions in CV and CG has nourished DL4SciVis research. 

Nevertheless, unlike images and videos in CV, SciVis data often requires the creation of visual representations and an explicit rendering process of those representations for display. 
For scalar fields, we either extract isosurfaces for visualization or map voxel values to colors and opacities via the transfer function for direct volume rendering. 
For vector fields, we place seeding points or curves in the domain to trace integral lines or surfaces for visualization. 
These resulting curves and surfaces share similarities with geometric models in CG. 
However, CG models are usually closed, while isosurfaces and flow surfaces are often non-closed. 
Besides geometric properties (e.g., curvature and torsion), flow surfaces also carry physical properties (e.g., density, viscosity, and tension) that need to be considered. 
Such differences among SciVis, CV, and CG often require customized DL solutions for best solving SciVis problems. 

In this paper, we present DL4SciVis, a state-of-the-art survey on DL works for SciVis. 
Our aims are three-fold: 
(1) introducing researchers to the recent advances in DL4SciVis; 
(2) categorizing DL4SciVis works in terms of domain setting, research task, learning type, network architecture, loss function, and evaluation metric; and 
(3) outlining research opportunities and open challenges. 
To our best knowledge, this paper is the first survey on DL4SciVis. 
We hope this comprehensive survey will help SciVis researchers understand this emerging direction and inspire them to join this vibrant research. 

\begin{table}[!]
\caption{All surveyed papers and their publication venues, tasks, and settings. 
The five tasks are data {\bf gen}eration [$\tikzcircle[black,fill=cherryblossompink]{2pt}$ $\tikzcircle[black,fill=brinkpink]{2pt}$ $\tikzcircle[black,fill=richcarmine]{2pt}$ $\tikzcircle[black,fill=carmine]{2pt}$], {\bf vis}ualization {\bf gen}eration [$\tikzcircle[black,fill=gold(web)(golden)]{2pt}$], prediction [$\tikzcircle[black,fill=green-yellow]{2pt}$ $\tikzcircle[black,fill=green(pigment)]{2pt}$], {\bf obj}ect {\bf det}ection and {\bf seg}mentation [$\tikzcircle[black,fill=azure(colorwheel)]{2pt}$], and {\bf feat}ure {\bf l}ea{\bf rn}ing and {\bf ext}raction [$\tikzcircle[black,fill=electricpurple]{2pt}$]. 
The four data generation subcategories are super-resolution [$\tikzcircle[black,fill=cherryblossompink]{2pt}$], compression and reconstruction [$\tikzcircle[black,fill=brinkpink]{2pt}$], translation [$\tikzcircle[black,fill=richcarmine]{2pt}$], and extrapolation [$\tikzcircle[black,fill=carmine]{2pt}$].  
The two prediction subcategories are data-relevant prediction [$\tikzcircle[black,fill=green-yellow]{2pt}$] and visualization-relevant prediction [$\tikzcircle[black,fill=green(pigment)]{2pt}$]. 
The five settings are scalar, vector, fluid {\bf sim}ulation, particle, and image.}
\vspace{-0.05in}
\label{tab:papers}
\centering
\setlength{\tabcolsep}{1pt}
\resizebox{\columnwidth}{!}{
\begin{tabular}{l|l|l|ccccc|ccccc}
\hline
paper & name & venue & \multicolumn{5}{c|}{task} & \multicolumn{5}{c}{setting} \\ \hline
          &           &            
          & \rotatebox{90}{data gen} & \rotatebox{90}{vis gen} & \rotatebox{90}{prediction} & \rotatebox{90}{obj det \& seg} & \rotatebox{90}{feat lrn \& ext  } 
          & \rotatebox{90}{scalar} & \rotatebox{90}{vector} & \rotatebox{90}{fluid sim} & \rotatebox{90}{particle} & \rotatebox{90}{image} 
           \\ \hline\hline
Zhou et al.\ \cite{Zhou-CGI17} & & CGI 
& $\tikzcircle[black,fill=cherryblossompink]{2pt}$ & & & & 
& $\checkmark$ & & & & \\       
Han and Wang~\cite{Han-VIS19} & TSR-TVD & TVCG
& $\tikzcircle[black,fill=cherryblossompink]{2pt}$ & & & & 
& $\checkmark$ & & & & \\       
Han and Wang~\cite{Han-TVCG} & SSR-TVD & TVCG
& $\tikzcircle[black,fill=cherryblossompink]{2pt}$ & & & & 
& $\checkmark$ & & & &  \\
Han et al.\ \cite{Han-VIS21} & STNet & TVCG 
& $\tikzcircle[black,fill=cherryblossompink]{2pt}$ & & & & 
& $\checkmark$ & & & &  \\
Wurster et al.\ \cite{Wurster-arXiv21} & & arXiv 
& $\tikzcircle[black,fill=cherryblossompink]{2pt}$ & & & & 
& $\checkmark$ & & & &  \\ 
Guo et al.\ \cite{Guo-PVIS20} & SSR-VFD & PVIS 
& $\tikzcircle[black,fill=cherryblossompink]{2pt}$ & & & & 
& & $\checkmark$ & & &  \\
Jakob et al.\ \cite{Jakob-VIS20} & & TVCG 
& $\tikzcircle[black,fill=cherryblossompink]{2pt}$ & & & & 
& & $\checkmark$ & & &  \\ 
Sahoo and Berger~\cite{Sahoo-EVISSP21} & IA-VFS & EVIS
& $\tikzcircle[black,fill=cherryblossompink]{2pt}$ & & & & 
& & $\checkmark$ & & &  \\
An et al.\ \cite{An-CGA21} & STSRNet & CG\&A 
& $\tikzcircle[black,fill=cherryblossompink]{2pt}$ & & & & 
& & $\checkmark$ & & &  \\ 
Han and Wang~\cite{Han-CG22} & TSR-VFD & C\&G 
& $\tikzcircle[black,fill=cherryblossompink]{2pt}$ & & & & 
& & $\checkmark$ & & &  \\ 
Xie et al.\ \cite{Xie-TOG18} & tempoGAN & TOG 
& $\tikzcircle[black,fill=cherryblossompink]{2pt}$ & & & & 
& & & $\checkmark$ & &  \\
Werhahn et al.\ \cite{Werhahn-CGIT19} & & CGIT 
& $\tikzcircle[black,fill=cherryblossompink]{2pt}$ & & & & 
& & & $\checkmark$ & &  \\ \hline
Wang et al.\ \cite{Wang-VIS19} & DeepOrganNet & TVCG 
& $\tikzcircle[black,fill=brinkpink]{2pt}$ & & & & 
& $\checkmark$ & & & &  \\ 
Lu et al.\ \cite{Lu-CGF21} & neurcomp & CGF 
& $\tikzcircle[black,fill=brinkpink]{2pt}$ & & & & 
& $\checkmark$ & & & &  \\ 
Weiss et al.\ \cite{Weiss-arXiv21-2} & fV-SRN & arXiv 
& $\tikzcircle[black,fill=brinkpink]{2pt}$ & & & & 
& $\checkmark$ & & & &  \\ 
Shi et al.\ \cite{Shi-PVIS22} & GNN-Surrogate & TVCG
& $\tikzcircle[black,fill=brinkpink]{2pt}$ & & & & 
& $\checkmark$ & & & &  \\ 
Han and Wang~\cite{Han-VI22} & VCNet & VI 
& $\tikzcircle[black,fill=brinkpink]{2pt}$ & & & & 
& $\checkmark$ & & & &  \\ 
Liu et al.\ \cite{Liu-JOV19} & & JOV 
& $\tikzcircle[black,fill=brinkpink]{2pt}$ & & & & 
& & $\checkmark$ & & &  \\
Han et al.\ \cite{Han-CGA19} & & CG\&A 
& $\tikzcircle[black,fill=brinkpink]{2pt}$ & & & & 
& & $\checkmark$ & & &  \\
Gu et al.\ \cite{Gu-CGA21} & VFR-UFD & CG\&A 
& $\tikzcircle[black,fill=brinkpink]{2pt}$ & & & & 
& & $\checkmark$ & & &  \\ \hline
Han et al.\ \cite{Han-VIS20} & V2V & TVCG 
& $\tikzcircle[black,fill=richcarmine]{2pt}$ & & & & 
& $\checkmark$ & & & &  \\
Gu et al.\ \cite{Gu-PVIS22} & Scalar2Vec & PVIS 
& $\tikzcircle[black,fill=richcarmine]{2pt}$ & & & & 
& $\checkmark$ & $\checkmark$ & & &  \\
Kim et al.\ \cite{Kim-CGF19-1} & Deep Fluids & CGF 
& $\tikzcircle[black,fill=richcarmine]{2pt}$ & & & & 
& & & $\checkmark$ & &  \\ 
Chu et al.\ \cite{Chu-TOG21} & & TOG 
& $\tikzcircle[black,fill=richcarmine]{2pt}$ & & & & 
& & & $\checkmark$ & &  \\ \hline
Wiewel et al.\ \cite{Wiewel-CGF19} & LSP & CGF
& $\tikzcircle[black,fill=carmine]{2pt}$ & & & & 
& & & $\checkmark$ & &  \\
Wiewel et al.\ \cite{Wiewel-CGF20} & LSS & CGF
& $\tikzcircle[black,fill=carmine]{2pt}$ & & & & 
& & & $\checkmark$ & &  \\ \hline\hline
Berger et al.\ \cite{Berger-TVCG19} & & TVCG 
& & $\tikzcircle[black,fill=gold(web)(golden)]{2pt}$ & & & 
& $\checkmark$ & & & &  \\
Hong et al.\ \cite{Hong-PVIS19} & DNN-VolVis & PVIS 
& & $\tikzcircle[black,fill=gold(web)(golden)]{2pt}$ & & & 
& $\checkmark$ & & & &  \\
He et al.\ \cite{He-VIS19} & InSituNet & TVCG 
& & $\tikzcircle[black,fill=gold(web)(golden)]{2pt}$ & & & 
& $\checkmark$ & & & &  \\
Weiss et al.\ \cite{Weiss-TVCG21} & & TVCG 
& & $\tikzcircle[black,fill=gold(web)(golden)]{2pt}$ & & & 
& $\checkmark$ & & & & \\
Weiss et al.\ \cite{Weiss-TVCG} & & TVCG 
& & $\tikzcircle[black,fill=gold(web)(golden)]{2pt}$ & & & 
& $\checkmark$ & & & &  \\ 
Weiss and Navab~\cite{Weiss-arXiv21-1} & DeepDVR & arXiv 
& & $\tikzcircle[black,fill=gold(web)(golden)]{2pt}$ & & & 
& $\checkmark$ & & & &  \\ \hline\hline
He et al.\ \cite{He-VI20} & CECAV-DNN & VI 
& & & $\tikzcircle[black,fill=green-yellow]{2pt}$ & & 
& $\checkmark$ & & & &  \\
Tkachev et al.\ \cite{Tkachev-TVCG21} & & TVCG 
& & & $\tikzcircle[black,fill=green-yellow]{2pt}$ & & 
& $\checkmark$ & & & &  \\
Hong et al.\ \cite{Hong-PVIS18} & & PVIS 
& & & $\tikzcircle[black,fill=green-yellow]{2pt}$ & & 
& & $\checkmark$ & & &  \\ 
Kim and G{\"u}nther~\cite{Kim-CGF19-2} & & CGF 
& & & $\tikzcircle[black,fill=green-yellow]{2pt}$ & & 
& & $\checkmark$ & & &  \\
Han et al.\ \cite{Han-arXiv21} & & arXiv 
& & & $\tikzcircle[black,fill=green-yellow]{2pt}$ & & 
& & $\checkmark$ & & &  \\ \hline
Yang et al.\ \cite{Yang-JOV19} & & JOV
& & & $\tikzcircle[black,fill=green(pigment)]{2pt}$ & & 
& $\checkmark$ & & & &  \\
Shi and Tao~\cite{Shi-TIST19} & & TIST
& & & $\tikzcircle[black,fill=green(pigment)]{2pt}$ & & 
& $\checkmark$ & & & &  \\
Engel and Ropinski~\cite{Engel-VIS20} & DVAO & TVCG
& & & $\tikzcircle[black,fill=green(pigment)]{2pt}$ & & 
& $\checkmark$ & & & &  \\ \hline\hline
Wang et al.\ \cite{Wang-VIS20} & VC-Net & TVCG
& & & & $\tikzcircle[black,fill=azure(colorwheel)]{2pt}$ &  
& $\checkmark$ & & & &  \\
Nguyen et al.\ \cite{Nguyen-arXiv21} & & arXiv
& & & & $\tikzcircle[black,fill=azure(colorwheel)]{2pt}$ &  
& $\checkmark$ & & & &  \\
Ghahremani et al.\ \cite{Ghahremani-TVCG} & NeuroConstruct & TVCG
& & & & $\tikzcircle[black,fill=azure(colorwheel)]{2pt}$ &  
& $\checkmark$ & & & &  \\
He et al.\ \cite{He-JOV22-2} & & JOV
& & & & $\tikzcircle[black,fill=azure(colorwheel)]{2pt}$ &  
& $\checkmark$ & & & & \\ 
Deng et al.\ \cite{Deng-JOV19} & Vortex-Net & JOV 
& & & & $\tikzcircle[black,fill=azure(colorwheel)]{2pt}$ &  
& & $\checkmark$ & & &  \\ 
Berenjkoub et al.\ \cite{Berenjkoub-VISSP20} & & VIS 
& & & & $\tikzcircle[black,fill=azure(colorwheel)]{2pt}$ &  
& & $\checkmark$ & & &  \\
Kashir et al.\ \cite{Kashir-JOV21} & & JOV 
& & & & $\tikzcircle[black,fill=azure(colorwheel)]{2pt}$ &  
& & $\checkmark$ & & &  \\
Borkiewicz et al.\ \cite{Borkiewicz-VISSP21} & CloudFindr & VIS
& & & & $\tikzcircle[black,fill=azure(colorwheel)]{2pt}$ &  
& & & & & $\checkmark$ \\  \hline\hline
Raji et al.\ \cite{Raji-EGPGV17} & & EGPGV 
& & & & & $\tikzcircle[black,fill=electricpurple]{2pt}$  
& $\checkmark$ & & & &  \\
Cheng et al.\ \cite{Cheng-TVCG19} & & TVCG 
& & & & & $\tikzcircle[black,fill=electricpurple]{2pt}$  
& $\checkmark$ & & & &  \\
Porter et al.\ \cite{Porter-VISSP19} & & VIS 
& & & & & $\tikzcircle[black,fill=electricpurple]{2pt}$  
& $\checkmark$ & & & &  \\
Tkachev et al.\ \cite{Tkachev-TVCG} & S4 & TVCG 
& & & & & $\tikzcircle[black,fill=electricpurple]{2pt}$  
& $\checkmark$ & & & &  \\
He et al.\ \cite{He-JOV22-1} & ScalarGCN & JOV
& & & & & $\tikzcircle[black,fill=electricpurple]{2pt}$  
& $\checkmark$ & & & &  \\
Han et al.\ \cite{Han-TVCG20} & FlowNet & TVCG 
& & & & & $\tikzcircle[black,fill=electricpurple]{2pt}$  
& & $\checkmark$ & & &  \\
Han and Wang~\cite{Han-CGF22} & SurfNet & CGF
& & & & & $\tikzcircle[black,fill=electricpurple]{2pt}$  
& $\checkmark$ & $\checkmark$ & & &  \\
Chu and Thuerey~\cite{Chu-TOG17} & & TOG 
& & & & & $\tikzcircle[black,fill=electricpurple]{2pt}$  
& & & $\checkmark$ & &  \\ 
Liu et al.\ \cite{Liu-arXiv19} & & arXiv 
& & & & & $\tikzcircle[black,fill=electricpurple]{2pt}$  
& & & & $\checkmark$ &  \\
Li and Shen~\cite{Li-TVCG} & & TVCG 
& & & & & $\tikzcircle[black,fill=electricpurple]{2pt}$  
& & & & $\checkmark$ &  \\
Zhu et al.\ \cite{Zhu-CGA21} & & CG\&A 
& & & & & $\tikzcircle[black,fill=electricpurple]{2pt}$  
& & & & & $\checkmark$  \\ \hline
\end{tabular}
}
\end{table}

\vspace{-0.1in}
\section{Scope of Survey}
\vspace{-0.025in}

To collect the related papers for this survey, we started with familiar papers published in VIS-relevant venues and searched the DBLP website.  
We scanned six visualization journals ({\em IEEE Transactions on Visualization and Computer Graphics}, {\em Computer Graphics Forum}, {\em IEEE Computer Graphics and Applications}, {\em Computers \& Graphics}, {\em Journal of Visualization}, and {\em Visual Informatics}). 
In addition, we searched three visualization conferences (VIS, EuroVis, and PacificVis) for relevant full (PacificVis) and short papers (all three). 
The following keywords were used to identify related papers in their titles: ``deep'', ``learning'', ``neural'', ``convolution'', ``generative'', ``adversarial'', ``recurrent'', ``CNN'', ``RNN'', and ``GAN''. 
We then scanned each paper's abstract, introduction, and conclusion to make a final decision. 

The criterion for selecting a paper is that the work utilizes deep neural networks (DNNs) to solve SciVis problems. In many cases, DL techniques are applied to address the end goal. However, we also included works that leverage DL techniques to solve intermediate steps that set up the solution to accomplish the final objective. 
To stay focused, we excluded ML but not DL works (e.g., using the random forest for data classification, support vector machine for data segmentation, or dictionary learning for data synthesis) and papers primarily related to VA or InfoVis instead of SciVis. 
We omitted works more closely related to CV or CG (e.g., image, video, point cloud, mesh geometry, illumination, shape, and motion) than SciVis topics. 
We augmented this collection by adding the more recent works of active DL4SciVis researchers from their websites and open-access repositories such as arXiv. 
Finally, we included a particular set of fluid flow simulation works. They are related to flow visualization (a key SciVis topic) and have inspired DL4SciVis researchers. 

It is often difficult to decide whether a paper is closer to SciVis than CV or CG. Ultimately, we check if the paper meets one of the two conditions for possible inclusion: (1) does the publication appear in a VIS venue? (2) has the publication significantly influenced SciVis work? Examples that meet each of the conditions are \cite{Wang-VIS20} and \cite{Xie-TOG18}, respectively. 

In the end, we gathered 59 papers for this survey, as listed in Table~\ref{tab:papers}. In the table, we order papers according to task (primary) and setting (secondary), followed by publication year within the same task-setting group. 
We note that this collection is by no means exhaustive. Instead, it serves as a representative one that allows us to perform a comprehensive review. 

\vspace{-0.1in}
\section{DL4SciVis Works}
\vspace{-0.025in}

This section discusses DL4SciVis works along six dimensions: domain setting, research task, learning type, network architecture, loss function, and evaluation metric. 
The discussion on the research task dimension provides the primary thread, allowing readers to dive into the surveyed works for in-depth comprehension. 
We also refer to their research tasks when categorizing the surveyed papers in the respective tables according to learning type, network architecture, loss function, and evaluation metric. 
The description of the other five dimensions aims to summarize and categorize these works from crucial aspects of design and evaluation for a comprehensive understanding.
In Tables~\ref{tab:terms-na}~and~\ref{tab:terms-lf-em}, we list acronyms that are used throughout this paper, where the indents represent hierarchical relationships.  

\begin{table}[t]
\caption{Acronyms for network architectures. Refer to Section~\ref{subsec:na} for a detailed explanation of basic network architectures and structures (i.e., AE, CNN, GAN, GNN, MLP, RNN).}
\vspace{-0.05in}
\label{tab:terms-na}
\centering
\setlength{\tabcolsep}{1pt}
\resizebox{\columnwidth}{!}{
\begin{tabular}{l|l}
\hline
 acronym & full name \\ \hline
AE~\cite{Ballard-AAAI87} & autoencoder\\ 
CNN~\cite{Lecun-IEEE98} & convolutional neural network\\
\hspace{3mm}DenseNet~\cite{Huang-CVPR17} & densely connected convolutional network\\
\hspace{3mm}EnhanceNet~\cite{Sajjadi-ICCV17} & enhance neural network\\
\hspace{3mm}ESPCN~\cite{Shi-CVPR16} & efficient sub-pixel convolutional network\\
\hspace{3mm}FCN~\cite{Long-CVPR15} & fully convolutional network\\
\hspace{3mm}FRVSR-Net~\cite{Sajjadi-CVPR18} & frame-recurrent video super-resolution network\\
\hspace{3mm}Geo-CNN~\cite{Lan-CVPR19} & geometric-induced CNN\\
\hspace{3mm}ResNet~\cite{He-CVPR16} & residual neural network\\
\hspace{3mm}Siamese~\cite{Bromley-NIPS93} & Siamese neural network (max-pooling CNN)\\
GAN~\cite{Goodfellow-NIPS14} & generative adversarial network\\
\hspace{3mm}BEGAN~\cite{Berthelot-arXiv17} & boundary equilibrium GAN\\
\hspace{6mm}cBEGAN~\cite{Marzouk-IJCNN19} & conditional BEGAN\\
\hspace{3mm}cGAN~\cite{Mirza-arXiv14} & conditional GAN\\
\hspace{3mm}ESRGAN~\cite{Wang-ECCVW18} & enhanced super-resolution GAN\\
\hspace{3mm}WGAN~\cite{Arjovsky-ICML17} & Wasserstein GAN\\
GNN~\cite{Gori-IJCNN05} & graph neural network\\
\hspace{3mm}GAE~\cite{Kipf-arXiv16} & graph autoencoder\\
\hspace{3mm}GCN~\cite{Kipf-ICLR17} & graph convolutional network\\
\hspace{3mm}GRN~\cite{Scarselli-TNN09} & graph recurrent network\\
\hspace{3mm}STGNN~\cite{Yu-IJCAI18} & spatial-temporal GNN\\
MLP~\cite{Rosenblatt-TR61} & multilayer perceptron\\
\hspace{3mm}FCCNN~\cite{Kawato-BC90} & fully connected cascade neural network\\
RNN~\cite{Rumelhart-TR85} & recurrent neural network\\
\hspace{3mm}LSTM~\cite{Hochreiter-NC97} & long short-term memory\\ 
\hspace{6mm}ConvLSTM~\cite{Shi-NIPS15} & convolutional LSTM\\
\hline 
\end{tabular}
}
\end{table}

\begin{table}[t]
\caption{Acronyms for loss functions and evaluation metrics.}
\vspace{-0.05in}
\label{tab:terms-lf-em}
\centering
\setlength{\tabcolsep}{1pt}
{\footnotesize
\begin{tabular}{l|l}
\hline
acronym & full name  (a.k.a. or closely similar name) \\ \hline
AAD & average angle difference\\
ACPPE & average critical point position error\\
AEDR~\cite{Ren-IJGI20} & adaptive edit distance on real sequence\\	
AER & average error rate\\
ALP & average last position\\
AWL~\cite{Wang-CVPR19} & adaptive wing loss\\
CD$^1$~\cite{Barrow-IJCAI77} & chamfer distance\\
CD$^2$ & cosine distance\\
& (cosine dissimilarity, cosine similarity)\\
CE & cross-entropy, cross-entropy loss\\
& (logarithmic loss, logistic loss, log loss)\\
\hspace{3mm}BCE & binary CE\\
CR & compression rate, compression ratio\\
& (reduction rate)\\
EMD~\cite{Rubner-IJCV00} & earth mover's distance\\
& (Wasserstein distance, Wasserstein metric)\\
F-score & F-1 score\\
& (Dice similarity coefficient, S\o{}rensen-Dice coefficient)\\
FID~\cite{Heusel-NIPS17} & Fr\'{e}chet inception distance\\
FN & false negative\\
FP & false positive\\
\hspace{3mm}FPR & FP rate\\
& (fall-out)\\
GSAL~\cite{Shi-TIST19} & geometric structure-aware loss\\
HD & Hausdorff distance\\
HR  & hit ratio\\
IOU & intersection over union\\
IS~\cite{Bruckner-CGF10} & isosurface similarity\\
Jaccard & Jaccard index in binary classification\\
& (visual object class)\\
LPIPS~\cite{Zhang-CVPR18} & learned perceptual image patch similarity\\
LSiM~\cite{Kohl-ICML20} & learned simulation metric\\
MAE & mean absolute error\\
& (L1 error, L1 loss, L1 norm)\\
\hspace{3mm}RAE & root absolute error\\
MCPD~\cite{Corouge-ISBI04} & mean of the closest point distances\\
ME & mean error\\
MI & mutual information\\
MOS & mean opinion score\\
MSE & mean squared error\\
& (L2 error, L2 loss, L2 norm)\\
\hspace{3mm}RMSE & root MSE\\
PPV & precision\\
& (positive predictive value)\\
PSNR & peak signal-to-noise ratio\\
REC~\cite{Bi-ICML03} & regression error characteristic\\
ROR & recall over rank\\
SC & silhouette coefficient\\
& (silhouette score)\\
SSIM~\cite{Wang-TIP04} & structural similarity index\\
& (structural dissimilarity index)\\
TN & true negative\\
TP & true positive\\
TPD & time partial derivative\\
TPR & true positive rate\\
& (sensitivity, recall, hit rate)\\
\hline 
\end{tabular}
}
\end{table}

\vspace{-0.1in}
\subsection{Domain Settings}
\vspace{-0.025in}

We group the surveyed works based on their {\em original} forms of data in their domain settings. The five settings are {\em scalar}, {\em vector}, {\em fluid simulation}, {\em particle}, and {\em image}. Note that the original data in such a setting is not necessarily the input to the DL model. For example, Berger et al.\ \cite{Berger-TVCG19} designed a GAN model for volume rendering, where the domain setting is a volumetric scalar field, and the input to the DL model is the viewpoint and transfer function. 

Table~\ref{tab:papers} shows that more than half of the papers handle scalar field data, including a single volume (e.g.,~\cite{Zhou-CGI17}), time-varying (e.g.,~\cite{Han-VIS19}), and multivariate (e.g.,~\cite{Han-VIS20}) volumetric data. This is not surprising as scalar field data are most commonly produced and widely available in SciVis. Apart from scalar field data, more than a quarter of the papers tackle vector field data, include 2D and 3D steady (e.g.,~\cite{Berenjkoub-VISSP20,Han-CGA19}) and unsteady (e.g.,~\cite{Kim-CGF19-2,Gu-CGA21}) vector fields. Two papers (i.e.,~\cite{Gu-PVIS22,Han-CGF22}) cover scalar and vector domains. We single out seven works (i.e.,~\cite{Xie-TOG18,Werhahn-CGIT19,Kim-CGF19-1,Chu-TOG21,Wiewel-CGF19,Wiewel-CGF20,Chu-TOG17}) and label them in the category of fluid simulation, as the primary focus of these works is simulation. Finally, two works (i.e.,~\cite{Liu-arXiv19,Li-TVCG}) target particle data, and the remaining two (i.e.,~\cite{Borkiewicz-VISSP21,Zhu-CGA21}) deal with image data. 

\begin{table*}[t]
\caption{Neural networks' inputs and outputs for data generation [$\tikzcircle[black,fill=cherryblossompink]{2pt}$ $\tikzcircle[black,fill=brinkpink]{2pt}$ $\tikzcircle[black,fill=richcarmine]{2pt}$ $\tikzcircle[black,fill=carmine]{2pt}$] papers.}
\vspace{-0.05in}
\label{tab:papers-IO-1}
\centering
\setlength{\tabcolsep}{3pt}
{\footnotesize
\begin{tabular}{l|l|l|l}
\hline
paper & name & input & output \\ \hline
Zhou et al.\ \cite{Zhou-CGI17} & & low-resolution volume & super-resolution volume \\       
Han and Wang~\cite{Han-VIS19} & TSR-TVD & two end volumes & intermediate volumes \\       
Han and Wang~\cite{Han-TVCG} & SSR-TVD & low-resolution volume & super-resolution volume \\
Han et al.\ \cite{Han-VIS21} & STNet & low-resolution two end volumes & super-resolution intermediate volumes \\
Wurster et al.\ \cite{Wurster-arXiv21} & & low-resolution volume & multi-resolution super-resolution volume \\ 
Guo et al.\ \cite{Guo-PVIS20} & SSR-VFD & low-resolution field & super-resolution field \\
Jakob et al.\ \cite{Jakob-VIS20} & & low-resolution flow map & super-resolution flow map \\ 
Sahoo and Berger~\cite{Sahoo-EVISSP21} & IA-VFS & low-resolution field & super-resolution field \\
An et al.\ \cite{An-CGA21} & STSRNet & low-resolution two end fields & super-resolution intermediate fields \\ 
Han and Wang~\cite{Han-CG22} & TSR-VFD & two end volumes & intermediate volumes \\ 
Xie et al.\ \cite{Xie-TOG18} & tempoGAN & low-resolution volume & super-resolution volume \\
Werhahn et al.\ \cite{Werhahn-CGIT19} & & low-resolution volume & super-resolution volume \\ \hline
Wang et al.\ \cite{Wang-VIS19} & DeepOrganNet & 3D/4D-CT projection or X-ray image & 3D/4D lung model \\ 
Lu et al.\ \cite{Lu-CGF21} & neurcomp & voxel coordinate & scalar value \\ 
Weiss et al.\ \cite{Weiss-arXiv21-2} & fV-SRN & voxel coordinate & density or color \\ 
Shi et al.\ \cite{Shi-PVIS22} & GNN-Surrogate & simulation parameters & output field with adaptive resolution \\ 
Han and Wang~\cite{Han-VI22} & VCNet & incomplete volume & complete volume \\
Liu et al.\ \cite{Liu-JOV19} & & volume patch & feature vector \\
Han et al.\ \cite{Han-CGA19} & & low-resolution field & high-resolution field \\
Gu et al.\ \cite{Gu-CGA21} & VFR-UFD & low-quality fields & high-quality fields \\ \hline
Han et al.\ \cite{Han-VIS20} & V2V & source variable & target variable \\
Gu et al.\ \cite{Gu-PVIS22} & Scalar2Vec & scalar field volume & vector field volume \\
Kim et al.\ \cite{Kim-CGF19-1} & Deep Fluids & velocity vector field, solver parameters & divergence-free velocity field \\ 
Chu et al.\ \cite{Chu-TOG21}& & density field & velocity field \\ \hline
Wiewel et al.\ \cite{Wiewel-CGF19} & LSP & multiple steps of pressure fields & future steps of pressure fields \\
Wiewel et al.\ \cite{Wiewel-CGF20} & LSS & early timesteps of simulation & future timesteps of simulation \\ \hline
\end{tabular}
}
\end{table*}

\vspace{-0.1in}
\subsection{Research Tasks}
\vspace{-0.025in}

Along the research task dimension, we classify the surveyed works into five categories: {\em data generation} [$\tikzcircle[black,fill=cherryblossompink]{2pt}$ $\tikzcircle[black,fill=brinkpink]{2pt}$ $\tikzcircle[black,fill=richcarmine]{2pt}$ $\tikzcircle[black,fill=carmine]{2pt}$], {\em visualization generation} [$\tikzcircle[black,fill=gold(web)(golden)]{2pt}$], {\em prediction} [$\tikzcircle[black,fill=green-yellow]{2pt}$ $\tikzcircle[black,fill=green(pigment)]{2pt}$], {\em objection detection and segmentation} [$\tikzcircle[black,fill=azure(colorwheel)]{2pt}$], and {\em feature learning and extraction} [$\tikzcircle[black,fill=electricpurple]{2pt}$], as shown in Table~\ref{tab:papers}. These research tasks are often the end goals of their respective works. However, in some cases, they only solve a subproblem, setting up a critical step to their final solution (e.g.,~\cite{Raji-EGPGV17,Cheng-TVCG19,Chu-TOG17,Zhu-CGA21}). 

Data generation tasks produce or reconstruct data or models from low-resolution versions (e.g.,~\cite{Zhou-CGI17}), feature representations (e.g.,~\cite{Han-CGA19}), or lower-dimensional counterparts (e.g.,~\cite{Wang-VIS19}) [$\tikzcircle[black,fill=cherryblossompink]{2pt}$ $\tikzcircle[black,fill=brinkpink]{2pt}$]. They also address data translation (e.g.,~\cite{Han-VIS20}) [$\tikzcircle[black,fill=richcarmine]{2pt}$] (e.g., translating from one variable sequence to another or from an input field to another that satisfies specific properties or constraints) and extrapolation (e.g.,~\cite{Wiewel-CGF19}) [$\tikzcircle[black,fill=carmine]{2pt}$] (e.g., generating historical or future data given the current data) issues.

Visualization generation tasks synthesize rendering results conditioned on the input, such as viewpoint (e.g.,~\cite{Hong-PVIS19}), transfer function (e.g.,~\cite{Berger-TVCG19}), and other parameters (e.g.,~\cite{He-VIS19}). Their goal is to produce unseen visualization results given the new input parameters without going through the traditional rendering pipelines or rerunning simulations under different parameter settings.  

Prediction tasks estimate class memberships, quality scores, future values, etc. The prediction results can assist users in making selections (e.g.,~\cite{Tkachev-TVCG21}), recommendations (e.g.,~\cite{Yang-JOV19}), or planning (e.g.,~\cite{Hong-PVIS18}) accordingly. These predictions can be data-relevant [$\tikzcircle[black,fill=green-yellow]{2pt}$] (e.g., voxel value) or visualization-relevant [$\tikzcircle[black,fill=green(pigment)]{2pt}$] (e.g., viewpoint quality). 

Objection detection and segmentation tasks are common in CV and biomedical imaging. The goal is to detect objects (e.g., animals, pedestrians, vehicles) from image or video data or segment different components (e.g., heart, lung, vessels) from biomedical data. Researchers also classify vortex boundaries~\cite{Berenjkoub-VISSP20} or vortical structures~\cite{Kashir-JOV21} from flow data. 

Finally, feature learning and extraction tasks learn data representations in the latent space (e.g.,~\cite{Han-TVCG20}) and extract domain-specific features (e.g.,~\cite{He-JOV22-1}). 
The learned representations can guide downstream tasks such as clustering, filtering, and representative selection. 

From Tables~\ref{tab:papers-IO-1} to~\ref{tab:papers-IO-5}, we list the neural networks' inputs and outputs of these papers (one for each category) following the same order as shown in Table~\ref{tab:papers}. The input and output refer to the inference stage whenever applicable. 
In the following, we describe these papers in detail. 

\vspace{-0.05in}
\subsubsection{Data Generation \textnormal{[$\tikzcircle[black,fill=cherryblossompink]{2pt}$ $\tikzcircle[black,fill=brinkpink]{2pt}$ $\tikzcircle[black,fill=richcarmine]{2pt}$ $\tikzcircle[black,fill=carmine]{2pt}$]}}

Under the category of data generation, we further group the related papers into four subcategories: {\em super-resolution}, {\em compression and reconstruction}, {\em translation}, and {\em extrapolation}.

{\bf Super-resolution \textnormal{[$\tikzcircle[black,fill=cherryblossompink]{2pt}$]}.}
Super-resolution is a class of techniques that aim to enhance or increase the image, video, or volumetric data resolution. In SciVis, the resolution encompasses the three spatial dimensions and the temporal dimension. Due to the limited storage space, generating super-resolution data from their low-resolution counterparts brings the immediate benefit of storage-saving via data reduction. This is because only the low-resolution data and the trained network model need to be stored to recover the super-resolution data. DL-based super-resolution techniques thus provide domain scientists an alternative to manage their simulation data cost-effectively. 

For scalar field data, the work of Zhou et al.\ \cite{Zhou-CGI17} is the first known one that utilizes a CNN for volume upscaling. Their CNN pipeline includes three stages: block extraction and feature representation, non-linear mapping, and reconstruction. They could upscale a single scalar volume by a factor of 2 (i.e., the super-resolution volume is 8$\times$ the size of the low-resolution input volume). 
To push the limit of data reduction, Wurster et al.\ \cite{Wurster-arXiv21} proposed a hierarchical super-resolution solution for volumetric data reduction. The resulting neural network hierarchy enables multi-resolution super-resolution generation at varying scaling factors (from 2$\times$ to 64$\times$). Finally, their octree-based data representation solution can upscale multi-resolution data to a uniform resolution while minimizing artifacts along block boundaries. 
Han and Wang considered time-varying volumetric data and presented TSR-TVD~\cite{Han-VIS19} and SSR-TVD~\cite{Han-TVCG} for generating temporal super-resolution (TSR) and spatial super-resolution (SSR). Both works leverage GANs for network training and consider temporal coherence. TSR-TVD can achieve a maximal interpolation step of 11, while SSR-TVD can upscale the volumes by a factor of 4. More recently, Han et al.\ \cite{Han-VIS21} introduced STNet, an end-to-end generative framework that achieves simultaneously spatiotemporal super-resolution (STSR) for time-varying volume data. They argued that straightforward concatenating SSR and TSR solutions does not lead to high-quality STSR volumes due to error propagation and presented STNet results superior to those of SSR+TSR. 

For vector field data, Guo et al.\ \cite{Guo-PVIS20} designed SSR-VFD to upscale a 3D vector field by a scaling factor of 4 or 8. Due to the possibly vast differences among vector components, they employed three neural nets to train individual components. These networks jointly generate spatially coherent super-resolution of vector field data. 
%
%
Sahoo and Berger~\cite{Sahoo-EVISSP21} presented IA-VFS, a solution for integration-aware vector field super-resolution. IA-VFS follows SSR-VFD~\cite{Guo-PVIS20} but considers integral streamlines in loss function to improve network optimization. 
Han and Wang~\cite{Han-CG22} developed TSR-VFD that generates intermediate vector fields from temporally sparsely sampled vector fields. TSR-VFD utilizes two networks: InterpolationNet and MaskNet, to learn different scales of the data and can achieve a maximal interpolation step of 9. 
An et al.\ \cite{An-CGA21} proposed STSRNet, a joint space-time super-resolution framework for vector field visualization. STSRNet includes two stages: the first one synthesizes intermediate frames given the two end frames at the low spatial resolution, and the second one upscales these intermediate frames to high spatial resolution. 
Jakob et al.\ \cite{Jakob-VIS20} adopted different versions of CNN to generate super-resolution flow maps, which are fundamental to Lagrangian transport analysis. They also provided the community with a large numerical 2D fluid flow dataset for ML. 

For fluid simulation, Xie et al.\ \cite{Xie-TOG18} introduced tempoGAN, a generative solution that produces temporally coherent volumetric super-resolution fluid flow. Their newly designed temporal discriminator can yield consistent and high-quality temporal results. Later works such as SSR-TVD~\cite{Han-TVCG} follow this design. 
Werhahn et al.\ \cite{Werhahn-CGIT19} presented a multi-pass GAN for generating fluid flow super-resolution. Their solution takes several orthogonal passes to decompose the space-time generative task so that each pass can solve an easily manageable inference problem.

{\bf Compression and reconstruction \textnormal{[$\tikzcircle[black,fill=brinkpink]{2pt}$]}.} 
Another subcategory of data generation tasks focuses explicitly on data compression and reconstruction. 
For data compression, 
%
Lu et al.\ \cite{Lu-CGF21} followed the SIREN~\cite{Sitzmann-NIPS20} design and introduced neurcomp, a coordinate-based MLP for compressive neural representations of scalar field volume data. Once learned, the network itself becomes the compressed representation of the underlying data. By quantizing network weights, neurcomp could achieve an impressive CR over 1,000$\times$ while preserving important volumetric features.
Weiss et al.\ \cite{Weiss-arXiv21-2} presented fV-SRN, which improves neurcomp by leveraging GPU tensor cores to integrate the reconstruction task into on-chip raytracing kernels. They also supported random access reconstruction at arbitrary granularity for temporal reconstruction tasks. 
Liu et al.\ \cite{Liu-JOV19} developed an in situ compression technique based on GAN for computational fluid dynamics (CFD) data. Compared with discrete wavelet transform, their solution achieves a speedup of over 3$\times$ in compression time while allowing a tradeoff between CR and reconstruction accuracy.

For data reconstruction, Wang et al.\ \cite{Wang-VIS19} designed DeepOrganNet that reconstructs 3D/4D lung models from single-view CT projections or X-ray images. DeepOrganNet can reconstruct manifold meshes of lung models in high quality and high fidelity, which all previous DL-based shape reconstruction approaches cannot.  	
Shi et al.\ \cite{Shi-PVIS22} developed GNN-Surrogate that reconstructs simulation outputs given simulation parameters as input. They generated graph hierarchies from unstructured grids and used the cutting policy to steer the representation of the simulation output with adaptive resolutions. 
Han and Wang~\cite{Han-VI22} designed VCNet for volume completion. The GAN-based neural network can synthesize missing subvolumes of various shapes (e.g., cuboid, cylinder, hyperboloid, sphere, tetrahedron, and ring) and sizes (up to 50\% of the entire volume). 
Han et al.\ \cite{Han-CGA19} addressed the issue of reconstructing vector field data from representative streamlines. Their solution encompasses two steps: initializing a low-resolution field based on the input streamlines and upscaling the low-resolution field using a CNN. 
Gu et al.\ \cite{Gu-CGA21} extended the work of Han et al.\ \cite{Han-CGA19} to reconstruct unsteady vector fields from their streamline representations via diffusion and DL-based denoising. Their solution captures temporal coherence by considering multiple consecutive timesteps at a time and preserves spatial coherence through streamline-based network optimization. 

{\bf Translation \textnormal{[$\tikzcircle[black,fill=richcarmine]{2pt}$]}.}
In SciVis, DL-based data translation was inspired by image colorization~\cite{Cheng-ICCV15} and image-to-image translation (e.g., Pix2Pix~\cite{Isola-CVPR17}, CycleGAN~\cite{Zhu-ICCV17}). 
Han et al.\ \cite{Han-VIS20} designed a DL solution for variable-to-variable (V2V) translation in the context of multivariate time-varying volumetric data. Their work first utilizes U-Net to learn features from variable sequences and identify suitable pairs for translation. Then, V2V leverages a GAN to achieve the translation (i.e., inferring the target variable sequence given the source variable sequence). 
Gu et al.\ \cite{Gu-PVIS22} presented Scalar2Vec that translates scalar fields to vector fields via DL. They followed the same approach as Han et al.\ \cite{Han-VIS20} to pick suitable scalar variables for the translation. The CNN-based network takes a set of sampled scalar field volumes as input and extracts their multi-scale information to synthesize the corresponding vector field volumes. 
Kim et al.\ \cite{Kim-CGF19-1} developed Deep Fluids for parameterized fluid simulations. Their generative model uses physics-informed loss functions to generate divergence-free velocity fields given the input velocity vectors and solver parameters. 
Chu et al.\ \cite{Chu-TOG21} aimed to infer velocity fields from density fields (i.e., translating density fields to velocity fields) via a data-driven cGAN model~\cite{Mirza-arXiv14}. Their work also provides multiple controls, such as physical parameters and kinetic energy, for fluid generation. 
We point out that some of these works (i.e.,~\cite{Han-VIS20,Chu-TOG21}) also imply data reduction. The trained neural nets can infer a previously unseen target variable sequence of later timesteps from the corresponding source sequence or a vector field from a scalar field, thus omitting the need to store the target variable or vector field. 

\begin{table*}[t]
\caption{Neural networks' inputs and outputs for visualization generation [$\tikzcircle[black,fill=gold(web)(golden)]{2pt}$] papers.}
\vspace{-0.05in}
\label{tab:papers-IO-2}
\centering
\setlength{\tabcolsep}{3pt}
{\footnotesize
\begin{tabular}{l|l|l|l}
\hline
paper & name & input & output \\ \hline
Berger et al.\ \cite{Berger-TVCG19} & & new viewpoint and transfer function & synthesized rendering conditioned on input \\
Hong et al.\ \cite{Hong-PVIS19} & DNN-VolVis & original rendering, goal effect, new viewpoint & synthesized rendering conditioned on input \\
He et al.\ \cite{He-VIS19} & InSituNet & ensemble simulation parameters & synthesized rendering conditioned on input \\
Weiss et al.\ \cite{Weiss-TVCG21} & & low-resolution isosurface maps, optical flow & high-resolution isosurface maps\\
Weiss et al.\ \cite{Weiss-TVCG} & & low-resolution image & high-resolution image \\ 
Weiss and Navab~\cite{Weiss-arXiv21-1} & DeepDVR & volume, viewpoint & rendering image \\ \hline
\end{tabular}
}
\end{table*}

\begin{table*}[t]
\caption{Neural networks' inputs and outputs for prediction [$\tikzcircle[black,fill=green-yellow]{2pt}$ $\tikzcircle[black,fill=green(pigment)]{2pt}$] papers.}
\vspace{-0.05in}
\label{tab:papers-IO-3}
\centering
\setlength{\tabcolsep}{3pt}
{\footnotesize
\begin{tabular}{l|l|l|l}
\hline
paper & name & input & output \\ \hline
He et al.\ \cite{He-VI20} & CECAV-DNN & sequence of ensemble pairs & likelihood each member from one ensemble \\
Tkachev et al.\ \cite{Tkachev-TVCG21} & & local spatiotemporal patch & future voxel value at patch center \\
Hong et al.\ \cite{Hong-PVIS18} & & movement sequence & probability vector of next movement \\ 
Kim and G{\"u}nther~\cite{Kim-CGF19-2} & & unsteady 2D vector field & reference frame transformation \\
Han et al.\ \cite{Han-arXiv21} & & particle start location, file cycles & particle end location \\ \hline
Yang et al.\ \cite{Yang-JOV19} & & volume rendering under viewpoint & viewpoint quality score \\
Shi and Tao~\cite{Shi-TIST19} & & volume rendering image & estimated viewpoint \\
Engel and Ropinski~\cite{Engel-VIS20} & DVAO & intensity volume, opacity volume or transfer function & AO volume \\ \hline
\end{tabular}
}
\end{table*}

{\bf Extrapolation \textnormal{[$\tikzcircle[black,fill=carmine]{2pt}$]}.}
Extrapolation aims to generate historical or future data values based on the current values. 
For fluid simulation, Wiewel et al.\ \cite{Wiewel-CGF19} presented latent space physics (LSP), an LSTM-CNN hybrid approach to predict the changes of pressure fields over time for fluid flow simulation. LSP can achieve 150$\times$ speedups compared with a regular pressure solver, a significant boost in simulation performance.
Wiewel et al.\ \cite{Wiewel-CGF20} proposed latent space subdivision (LSS), an end-to-end DL-solution for robust prediction future timesteps of complex fluid simulations with high temporal stability. Using CNN and stacked LSTM, LSS achieves both spatial compression and temporal prediction. 

\vspace{-0.05in}
\subsubsection{Visualization Generation \textnormal{[$\tikzcircle[black,fill=gold(web)(golden)]{2pt}$]}}

All existing DL4SciVis works in visualization generation are related to either direct volume rendering or isosurface rendering. 
Berger et al.\ \cite{Berger-TVCG19} presented a generative model for volume rendering where a GAN was trained on a large collection of volume rendering images under different viewpoints and transfer functions. Once trained, the model can infer novel rendering conditioned on new viewpoints and transfer functions without following the traditional rendering pipeline. 
Hong et al.\ \cite{Hong-PVIS19} designed DNN-VolVis, a DNN for volume visualization. Their goal is to synthesize volume rendering results from the original input rendering under a given target effect and new viewing parameters. Thus, without knowing the underlying transfer function, their generative framework supports volume exploration in a reverse manner. 
Weiss and Navab~\cite{Weiss-arXiv21-1} trained DeepDVR, an end-to-end DNN that explicitly models the direct volume rendering process, including feature extraction, classification, and composition. Their solution generates similar direct volume rendering results from examples in the image space, eliminating the need for explicit feature design and manual transfer function specification. 

He et al.\ \cite{He-VIS19} developed InSituNet for parameter-space exploration of ensemble simulations. The training data (i.e., visualization images conditioned on visual mappings and view parameters) were collected in situ. Then, they trained a convolutional regression model offline that learns the mapping from simulation parameters to visualization outputs. The trained model supports interactive post hoc exploration and analysis by synthesizing images from novel parameter settings. 
Weiss et al.\ \cite{Weiss-TVCG21} generated super-resolution isosurface rendering images from their low-resolution counterparts using FRVSR-Net (a fully convolutional frame-recurrent neural network). The network takes low-resolution isosurface maps (mask, normal, and depth) and optical flow as input and outputs high-resolution isosurface maps, including mask, normal, depth, and ambient occlusion (AO) maps. 
Weiss et al.\ \cite{Weiss-TVCG} aimed to learn the correspondence between the data, sampling patterns, and generated images. To achieve the goal, they introduced an end-to-end neural rendering framework consisting of two networks (i.e., importance network and reconstruction network). The former infers the importance map from low-resolution rendering images, and the latter recovers high-resolution images from sparse samples.

\begin{table*}[t]
\caption{Neural networks' inputs and outputs for object detection and segmentation [$\tikzcircle[black,fill=azure(colorwheel)]{2pt}$] papers.}
\label{tab:papers-IO-4}
\vspace{-0.05in}
\centering
\setlength{\tabcolsep}{3pt}
{\footnotesize
\begin{tabular}{l|l|l|l}
\hline
paper & name & input & output \\ \hline
Wang et al.\ \cite{Wang-VIS20} & VC-Net & 3D volume patch, multislice composited 2D MIP & vessel mask \\
Nguyen et al.\ \cite{Nguyen-arXiv21} & & cryo-EM image, dense pseudo labels & soft labels \\
Ghahremani et al.\ \cite{Ghahremani-TVCG} & NeuroConstruct & batch of grayscale images & probability map \\
He et al.\ \cite{He-JOV22-2} &  & super-voxel graph with neighborhood relations & feature classification per super-voxel \\
Deng et al.\ \cite{Deng-JOV19} & Vortex-Net & sample local patch & hard labels \\ 
Berenjkoub et al.\ \cite{Berenjkoub-VISSP20} & & velocity patch & binary classification of vortex boundary \\
Kashir et al.\ \cite{Kashir-JOV21} & & input map (velocity, vorticity) & binary segmentation of vortical structure  \\
Borkiewicz et al.\ \cite{Borkiewicz-VISSP21} & CloudFindr & image patch & predicted mask \\ \hline
\end{tabular}
}
\end{table*}

\vspace{-0.05in}
\subsubsection{Prediction \textnormal{[$\tikzcircle[black,fill=green-yellow]{2pt}$ $\tikzcircle[black,fill=green(pigment)]{2pt}$]}}

Under the category of data prediction, we further group the related papers into two subcategories: {\em data-relevant prediction} and {\em visualization-relevant prediction}. 

{\bf Data-relevant prediction \textnormal{[$\tikzcircle[black,fill=green-yellow]{2pt}$]}.}
For scalar field data, He et al.\ \cite{He-VI20} designed CECAV-DNN that predicts ensemble similarity for collective ensemble comparison and visualization (CECAV). Given a sequence of ensemble pairs (each ensemble is a collection of scalar fields), they trained the DNN to assign a likelihood score to each scalar field, indicating the probability that the field is from one ensemble rather than the other. After training, three levels of comparison: dimensionality comparison, member comparison, and region comparison, are provided for ensemble comparison and visualization.
Tkachev et al.\ \cite{Tkachev-TVCG21} developed a local prediction model for spatiotemporal volume visualization. Their goal is to detect irregular processes (i.e., outliers) in the space-time data. To this end, they designed a neural network that takes local spatiotemporal patches and predicts future voxel values at patch centers. The predicted values' deviation from the ground-truth values suggests mispredicted spatiotemporal regions for further study. 

For vector field data, Hong et al.\ \cite{Hong-PVIS18} aimed to predict the access pattern for parallel particle tracing. Their LSTM-based model learns the access pattern from a small set of pathline samples. Such prediction results can assist workload balancing by prefetching data blocks to reduce I/O costs and improve time efficiency. 
Kim and G{\"u}nther~\cite{Kim-CGF19-2} designed a CNN to predict reference frame transformations for 2D unsteady vector fields. Their solution can tackle noisy inputs and data with resampling artifacts by performing filtering and reference frame extraction end-to-end. 
Han et al.\ \cite{Han-arXiv21} predicted particle end locations for Lagrangian-based particle tracing. Their MLP-based model learns particle end locations given their start locations and file cycles. The trained model can predict new particle trajectories with a small memory cost and fast inference. 

{\bf Visualization-relevant prediction \textnormal{[$\tikzcircle[black,fill=green(pigment)]{2pt}$]}.}
This subcategory of prediction tasks estimates visualization-related quality or parameters. 
Yang et al.\ \cite{Yang-JOV19} designed a CNN-based model to estimate the viewpoint quality given a volume rendering image. The aim is to mimic the traditional scoring method and user preference and predict viewpoint quality close to human judgment. 
Shi and Tao~\cite{Shi-TIST19} attempted to estimate the viewpoint given a volume rendering image. Their CNN-based viewpoint estimation framework features an overfit-resistant image rendering strategy for training data generation and a geometric structure-aware loss design. 
Engel and Ropinski~\cite{Engel-VIS20} presented DVAO, deep volumetric AO, that predicts the AO volume given the original intensity volume and opacity information specified in the form of opacity volume or transfer function descriptor. DVAO supports real-time volume interaction with per-voxel AO estimation. 

\begin{table*}[t]
\caption{Neural networks' inputs and outputs for feature learning and extraction [$\tikzcircle[black,fill=electricpurple]{2pt}$] papers.}
\label{tab:papers-IO-5}
\vspace{-0.05in}
\centering
\setlength{\tabcolsep}{3pt}
{\footnotesize
\begin{tabular}{l|l|l|l}
\hline
paper & name & input & output \\ \hline
Raji et al.\ \cite{Raji-EGPGV17} & & image pairs & feature vectors\\
Cheng et al.\ \cite{Cheng-TVCG19} & & volume patch & feature vector \\
Porter et al.\ \cite{Porter-VISSP19} & & volume & feature vector \\
Tkachev et al.\ \cite{Tkachev-TVCG} & S4 & local spatiotemporal patches & feature vectors \\
He et al.\ \cite{He-JOV22-1} & ScalarGCN & Scalar-Graph with local and global connections & feature vector per variable \\
Han et al.\ \cite{Han-TVCG20} & FlowNet & streamline or stream surface & feature vector \\
Han and Wang~\cite{Han-CGF22} & SurfNet & isosurface or stream surface & node features \\
Chu and Thuerey~\cite{Chu-TOG17} & & flow patch pairs & feature vectors \\ 
Liu et al.\ \cite{Liu-arXiv19} & & density field & feature vector \\
Li and Shen~\cite{Li-TVCG} & & particle patch & feature vector \\
Zhu et al.\ \cite{Zhu-CGA21} & & feature position in initial scatterplot & inferred feature position in new scatterplot \\ \hline
\end{tabular}
}
\end{table*}

\vspace{-0.05in}
\subsubsection{Object Detection and Segmentation \textnormal{[$\tikzcircle[black,fill=azure(colorwheel)]{2pt}$]}} 

Existing DL4SciVis works on object detection and segmentation heavily utilize U-Net~\cite{Ronneberger-MICCAI15}, initially designed for biomedical image segmentation. 
For scalar field data, 
Wang et al.\ \cite{Wang-VIS20} designed VC-Net, a deep volume composition network for segmenting vessels from highly sparse and noisy biomedical image data. Their paradigm includes a dual-stream component and the bi-directional operations between them. The 3D volume segmentation stream follows a 3D U-Net design, and the 2D composited maximum intensity projection (MIP) segmentation stream uses a half 2D U-Net. To achieve effective exploration, VC-Net combines direct 3D volume processing (3D stream) and volume-rendered clues (2D stream). 
Nguyen et al.\ \cite{Nguyen-arXiv21} presented a semi-supervised volume visualization solution for cryo-EM data. Their solution includes two segmentation algorithms (a weak one and a powerful DL-based one) to produce soft segmentation results guiding the transfer function design. They compared three models: 3D U-Net, 3D U-Net+ResNet, and 3D DenseNet, along with three losses: BCE, MSE, and AWL, and reported that 3D U-Net+ResNet with MSE loss works best. 
Ghahremani et al.\ \cite{Ghahremani-TVCG} developed NeuroConstruct to reconstruct 3D neurites from optical microscopy brain images. Their 3D CNN-based segmentation model consists of multiple stages of residual U-block (RSU) connected in the big U-structure.

For vector field and image data, 
Deng et al.\ \cite{Deng-JOV19} designed Vortex-Net, a CNN-based method for vortex identification. Their binary classification solution benefits from global and local vortex identification methods to achieve better performance (both speed and accuracy). 
Berenjkoub et al.\ \cite{Berenjkoub-VISSP20} presented a CNN to identify vortex boundary. They experimented with three CNN architectures: conventional CNN, ResNet, and U-Net, and reported that U-Net achieves the best performance in the binary classification task.
Kashir et al.\ \cite{Kashir-JOV21} utilized an FCN to identify vortical structures in 2D fluid flow. The model takes velocity and vorticity maps as input and produces pixel-wise semantic segmentation results. In addition, they investigated the symmetric U-shaped network structure to find the optimal settings to best extract vortical structures. 
Borkiewicz et al.\ \cite{Borkiewicz-VISSP21} developed CloudFindr to detect cloud from satellite image data. Their U-Net-based solution produces a predicted mask given the input image patch. 

Beyond U-Net, researchers also explored the use of GNN for volume classification. 
He et al.\ \cite{He-JOV22-2} generated a super-voxel graph from a scalar volumetric dataset where a node represents a super-voxel (i.e., a group of voxels with similar spatial locations and properties), and an edge represents the neighborhood relation between the corresponding super-voxels. They then utilized a GCN to learn node embedding. Finally, the output of the GCN goes through an MLP to predict the label of each node for volume classification. 

\vspace{-0.05in}
\subsubsection{Feature Learning and Extraction \textnormal{[$\tikzcircle[black,fill=electricpurple]{2pt}$]}}

For scalar field data, many solutions use CNN-based neural networks for feature learning and extraction. 
Raji et al.\ \cite{Raji-EGPGV17} trained a deep Siamese network 
to extract feature vectors from image pairs (real-world photographs and volume rendering images) and judge whether nor not the input images are similar. The goal is to optimize rendering parameters via an evolutionary process to match the features of rendering images with those in the photographs. 
Cheng et al.\ \cite{Cheng-TVCG19} applied a pre-trained CNN to learn voxel neighborhood information. They then employed vector quantization to the high-level features extracted from volume patches to generate the characteristic feature vector to support the hierarchical exploration of complex volumetric structures. 
Porter et al.\ \cite{Porter-VISSP19} leveraged an AE to encode each timestep of a time-varying volumetric dataset into a feature vector, which was then projected to an abstract 2D space for identifying representative timesteps. Their approach can naturally handle multivariate datasets using a multichannel input which previous works cannot. 
%
%
Tkachev et al.\ \cite{Tkachev-TVCG} designed S4, self-supervised learning of spatiotemporal similarity, for supporting explicit similarity queries of scientific datasets. They employed a Siamese network to extract feature vectors from local spatiotemporal patches and judge if they are from the same neighborhood. 

Beyond CNNs, researchers have also investigated GNN-based solutions for feature learning and extraction. 
He et al.\ \cite{He-JOV22-1} designed ScalarGCN, a GNN-based solution for scalar-value association analysis of volumes. ScalarGCN aims to learn the high-order topological structural relationships of multiple variables using a multilayer GCN with the self-attention mechanism. The input to ScalarGCN is Scalar-Graph, where nodes represent sampled scalar values from multivariate data, and edges encode local (within the same variable) and global (across different variables) connections. Node features consider context, spatial, and gradient distributions. ScalarGCN performs local learning for node embedding and global learning for variable embedding. 

For vector field data and fluid simulation, 
Han et al.\ \cite{Han-TVCG20} introduced FlowNet, which is an AE for learning the latent features of streamlines and stream surfaces implicitly. They compared three losses: BCE, Dice, and MSE, and reported that BCE achieves the best results. The learned features are projected into a low-dimensional space to support clustering, filtering, and selection of representatives. Their voxel-based representation makes the solution generally applicable to any 3D data or their visual representations (e.g.,~\cite{Porter-VISSP19}), albeit not necessarily efficient when the representations are sparse in the 3D domain. 
Han and Wang~\cite{Han-CGF22} designed SurfNet, a GCN-based solution for learning node and surface features for isosurfaces and stream surfaces. Compared with FlowNet~\cite{Han-TVCG20}, training SurfNet is 10$\times$ to 20$\times$ faster per epoch and inferring is 70$\times$ to 170$\times$ faster while the model reduction is 300$\times$ to 1,300$\times$. 
Chu and Thuerey~\cite{Chu-TOG17} employed two identical CNNs similar to the Siamese network to extract feature vectors from two flow patches: a coarse approximation and a refined version of two flow simulations of the same effect. Their end goal is to identify the best-matched flow patch from a fluid repository of pre-computed data to refine a new coarse input for volumetric synthesis.  

For particle and image data, 
Liu et al.\ \cite{Liu-arXiv19} investigated the use of a residual AE for feature learning of a particle dataset. Their goal is to achieve in situ data reduction that preserves features (i.e., gas bubbles in a fluid). 
Li and Shen~\cite{Li-TVCG} leveraged a Geo-CNN~\cite{Lan-CVPR19} to extract features from particle data that capture their spatial-physical attribute relationships without explicitly knowing their spatial connectivity information. The learned feature information was utilized in the subsequent feature tracking. 
Zhu et al.\ \cite{Zhu-CGA21} designed a cascade neural network that takes hyperspectral image features as input and infers a scatterplot where domain experts customized the cluster centers. Once trained, the same network can be used for studying time-varying hyperspectral images without retraining. 

\begin{table*}[t]
\caption{All surveyed papers and their learning types organized under the five research tasks. The $^\star$ sign indicates the work includes a pre-training step. The $^+$ sign indicates the work uses a pre-trained model.}
\vspace{-0.05in}
\label{tab:learning}
\centering
{\scriptsize
\begin{tabular}{l|c|c|c|c|c}
\hline
 & \multicolumn{5}{c}{task} \\ \cline{2-6}
learning & data gen [$\tikzcircle[black,fill=cherryblossompink]{2pt}$ $\tikzcircle[black,fill=brinkpink]{2pt}$ $\tikzcircle[black,fill=richcarmine]{2pt}$ $\tikzcircle[black,fill=carmine]{2pt}$] & vis gen [$\tikzcircle[black,fill=gold(web)(golden)]{2pt}$] & prediction [$\tikzcircle[black,fill=green-yellow]{2pt}$ $\tikzcircle[black,fill=green(pigment)]{2pt}$] & obj det \& seg [$\tikzcircle[black,fill=azure(colorwheel)]{2pt}$] & feat lrn \& ext [$\tikzcircle[black,fill=electricpurple]{2pt}$]\\ \hline  
\multirow{4}{*}{supervised} 
& 
$\tikzcircle[black,fill=cherryblossompink]{2pt}$\cite{An-CGA21}, 
$\tikzcircle[black,fill=richcarmine]{2pt}$\cite{Chu-TOG21}, 
$\tikzcircle[black,fill=brinkpink]{2pt}$\cite{Gu-CGA21},
$\tikzcircle[black,fill=richcarmine]{2pt}$\cite{Gu-PVIS22},
$\tikzcircle[black,fill=cherryblossompink]{2pt}$\cite{Guo-PVIS20},
$\tikzcircle[black,fill=brinkpink]{2pt}$\cite{Han-CGA19},
$\tikzcircle[black,fill=cherryblossompink]{2pt}$\cite{Han-TVCG}
& 
\cite{Berger-TVCG19}, \cite{He-VIS19} 
& 
$\tikzcircle[black,fill=green(pigment)]{2pt}$\cite{Engel-VIS20}, 
$\tikzcircle[black,fill=green-yellow]{2pt}$\cite{Han-arXiv21}
& 
\cite{Berenjkoub-VISSP20}, \cite{Borkiewicz-VISSP21} 
& 
\cite{Cheng-TVCG19}$^+$ 
\\
& 
$\tikzcircle[black,fill=cherryblossompink]{2pt}$\cite{Han-VIS19},
$\tikzcircle[black,fill=cherryblossompink]{2pt}$\cite{Han-CG22},
$\tikzcircle[black,fill=brinkpink]{2pt}$\cite{Han-VI22},
$\tikzcircle[black,fill=cherryblossompink]{2pt}$\cite{Han-VIS21}$^\star$,
$\tikzcircle[black,fill=richcarmine]{2pt}$\cite{Han-VIS20},
$\tikzcircle[black,fill=cherryblossompink]{2pt}$\cite{Jakob-VIS20},
$\tikzcircle[black,fill=richcarmine]{2pt}$\cite{Kim-CGF19-1}
& 
\cite{Hong-PVIS19}, \cite{Weiss-arXiv21-1}
& 
$\tikzcircle[black,fill=green-yellow]{2pt}$\cite{Hong-PVIS18},
$\tikzcircle[black,fill=green-yellow]{2pt}$\cite{Kim-CGF19-2}
& 
\cite{Deng-JOV19}, \cite{Ghahremani-TVCG} 
& \\
& 
$\tikzcircle[black,fill=cherryblossompink]{2pt}$\cite{Sahoo-EVISSP21},
$\tikzcircle[black,fill=brinkpink]{2pt}$\cite{Shi-PVIS22},
$\tikzcircle[black,fill=brinkpink]{2pt}$\cite{Wang-VIS19}$^+$, 
$\tikzcircle[black,fill=brinkpink]{2pt}$\cite{Weiss-arXiv21-2},
$\tikzcircle[black,fill=cherryblossompink]{2pt}$\cite{Werhahn-CGIT19}
& 
\cite{Weiss-TVCG21} 
& 
$\tikzcircle[black,fill=green(pigment)]{2pt}$\cite{Shi-TIST19}, 
$\tikzcircle[black,fill=green-yellow]{2pt}$\cite{Tkachev-TVCG21} 
& 
\cite{Kashir-JOV21}, \cite{Wang-VIS20}
&\\
& 
$\tikzcircle[black,fill=carmine]{2pt}$\cite{Wiewel-CGF19},
$\tikzcircle[black,fill=carmine]{2pt}$\cite{Wiewel-CGF20},
$\tikzcircle[black,fill=cherryblossompink]{2pt}$\cite{Wurster-arXiv21},
$\tikzcircle[black,fill=cherryblossompink]{2pt}$\cite{Xie-TOG18},  
$\tikzcircle[black,fill=cherryblossompink]{2pt}$\cite{Zhou-CGI17} 
& 
\cite{Weiss-TVCG}
& 
$\tikzcircle[black,fill=green(pigment)]{2pt}$\cite{Yang-JOV19}  
& &\\ \hline
\hspace{5mm}weakly-supervised & --- & --- & --- & --- & --- \\ \hline
semi-supervised & --- & --- & --- 
& 
\cite{He-JOV22-2} \cite{Nguyen-arXiv21}$^\star$
& 
\cite{Zhu-CGA21}
\\ \hline
unsupervised  & & & & & \\\hline
\multirow{2}{*}{\hspace{5mm}distributed} 
& 
$\tikzcircle[black,fill=brinkpink]{2pt}$\cite{Liu-JOV19}, 
$\tikzcircle[black,fill=brinkpink]{2pt}$\cite{Lu-CGF21}
& --- 
& 
$\tikzcircle[black,fill=green-yellow]{2pt}$\cite{He-VI20} 
& --- 
& 
\cite{Han-TVCG20}, \cite{Han-CGF22}, \cite{He-JOV22-1} 
\\
& & & & 
& 
 \cite{Li-TVCG}, \cite{Liu-arXiv19}, \cite{Porter-VISSP19} 
\\\hline
\hspace{5mm}disentangled & --- & --- & --- & --- & --- \\ \hline
\hspace{5mm}self-supervised 
& --- & --- & --- & --- 
& 
\cite{Chu-TOG17}, \cite{Raji-EGPGV17}, \cite{Tkachev-TVCG}  \\ \hline
\end{tabular}
}
\end{table*}

\vspace{-0.05in}
\subsubsection{Summary}

Comparing the five research tasks, we see many more works in data generation [$\tikzcircle[black,fill=cherryblossompink]{2pt}$ $\tikzcircle[black,fill=brinkpink]{2pt}$ $\tikzcircle[black,fill=richcarmine]{2pt}$ $\tikzcircle[black,fill=carmine]{2pt}$] than visualization generation [$\tikzcircle[black,fill=gold(web)(golden)]{2pt}$]. We reason that data generation tasks in SciVis share significant similarities with those in CV, making it relatively easier to work on even with the apparent challenge of handling 3D volume data instead of 2D image data. On the other hand, visualization generation tasks must consider different parameters (e.g., viewpoint, transfer function, ensemble parameters, etc.) and demand new solutions to assimilate such heterogeneous information into network design and training. Recent advances in CG, such as neural rendering~\cite{Tewari-CGF20} and differentiable rendering~\cite{Kato-arXiv20}, provide good opportunities for SciVis researchers to expand the current research in visualization generation. 
 
Within the category of data generation, there are more super-resolution [$\tikzcircle[black,fill=cherryblossompink]{2pt}$] works than compression and reconstruction [$\tikzcircle[black,fill=brinkpink]{2pt}$], translation [$\tikzcircle[black,fill=richcarmine]{2pt}$], and extrapolation [$\tikzcircle[black,fill=carmine]{2pt}$]. We can contribute this difference to the fact that data reconstruction may consider their visual representations as input, data translation adds extra complexity from multivariate relationships and multichannel variations, and data {\em extrapolation} is intrinsically more challenging than super-resolution (which can be treated as a form of data {\em interpolation}). Reconstruction, translation, and extrapolation could become a growth point for future research in data generation. 

Feature learning and extraction [$\tikzcircle[black,fill=electricpurple]{2pt}$] is a resounding theme for CV, CG, and VIS. In SciVis, feature definitions are usually application-specific, and in many cases, they are vague or even unknown. Therefore, explicitly or implicitly, learning features is the necessary first step toward effective analysis and visualization. By replacing manual feature engineering with automatic feature discovery, representation learning can help accomplish a wide variety of subsequent tasks critical to SciVis, such as dimensionality reduction, data clustering, representative selection, anomaly detection, data classification, and data generation. Due to the general need and the variety of data (scalar and vector, time-varying and multivariate) and their visual representations (line, surface, volume), we expect a strongly growing trend in DL-based solutions for feature or representation learning. 

Prediction tasks [$\tikzcircle[black,fill=green-yellow]{2pt}$ $\tikzcircle[black,fill=green(pigment)]{2pt}$] commonly serve as an intermediate step of a large problem which yields critical prediction results to assist downstream tasks. This makes unsupervised learning, particularly self-supervised learning, a suitable candidate for accomplishing such tasks. Thus, investigating the underexplored self-supervised learning solutions for making predictions or recommendations will certainly boost DL4SciVis research.

Finally, object detection and segmentation papers [$\tikzcircle[black,fill=azure(colorwheel)]{2pt}$] are often published in CV and biomedical imaging venues, even though medical visualization is a long-standing topic in SciVis. Nevertheless, we see that popular networks such as U-Net, initially designed for biomedical image segmentation, have been widely adopted and utilized for solving SciVis problems. 

\begin{figure}[t]
\begin{center}
\includegraphics[width=1.0\linewidth]{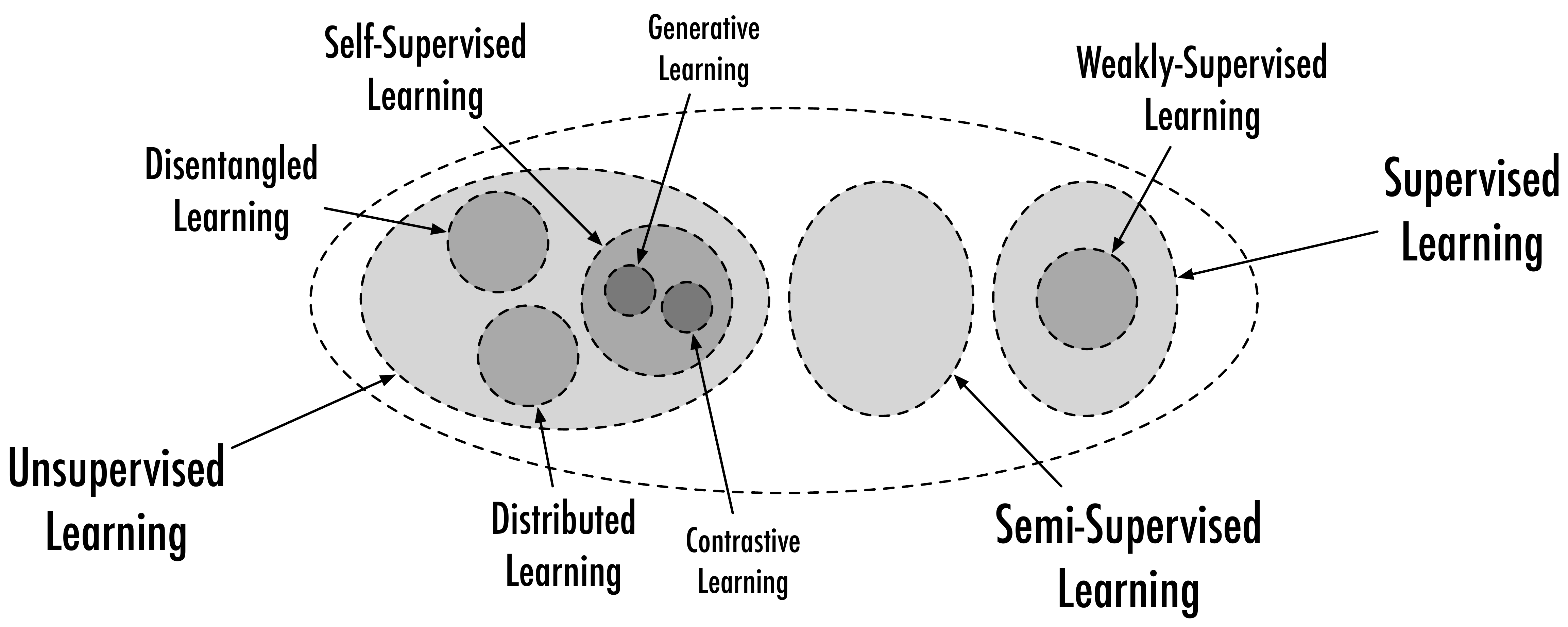}
\end{center}
\vspace{-0.1in}
\caption{The relationships among different learning types.}
\label{fig:learning}
\end{figure}

\vspace{-0.1in}
\subsection{Learning Types} 
\vspace{-0.025in}

As illustrated in Figure~\ref{fig:learning}, depending on how much labeled data are used for neural network training, DL tasks can be {\em supervised} (full labels), {\em semi-supervised} (partial labels), or {\em unsupervised} (no labels)~\cite{Jing-TPAMI21}. As a subset of supervised learning, {\em weakly-supervised} learning learns feature representations with coarse-grained or inaccurate labels. Under the umbrella of unsupervised learning, there are three types of learning: {\em distributed learning}, {\em disentangled learning}, and {\em self-supervised learning}. 

Distributed learning denotes the same data features across multiple {\em scalable} and {\em interdependent} layers, which are learned concurrently but non-linearly. Each layer describes the information with the same accuracy level while adjusted for the scale level. 

Unlike distributed learning, disentangled learning represents the data with {\em independent} features, each describing partial information, such as {\em content} and {\em style}. It is possible to learn disentangled representations from distributed representations by appropriate transformations. 

Self-supervised learning automatically generates the labels from the underlying data itself by leveraging its structure~\cite{Jing-TPAMI21}. This learning approach consists of two stages. The first stage trains unlabeled data in a {\em pretext task} (e.g., jigsaw puzzle) to generate representations. The second stage applies these representations to a {\em downstream task} (e.g., classification or segmentation) using only a small amount of labeled data.
Self-supervised learning can be {\em generative} or {\em contrastive}~\cite{Liu-TKDE}. 
Generative learning learns representations from data by fitting the data distribution. 
Contrastive learning aims at ``learning to compare'' through a noise contrastive estimation objective. 

Table~\ref{tab:learning} classifies all surveyed papers into different learning types. 
Overall, we can see that supervised learning is dominant across the four categories of research tasks: data generation [$\tikzcircle[black,fill=cherryblossompink]{2pt}$ $\tikzcircle[black,fill=brinkpink]{2pt}$ $\tikzcircle[black,fill=richcarmine]{2pt}$ $\tikzcircle[black,fill=carmine]{2pt}$], visualization generation [$\tikzcircle[black,fill=gold(web)(golden)]{2pt}$], prediction [$\tikzcircle[black,fill=green-yellow]{2pt}$ $\tikzcircle[black,fill=green(pigment)]{2pt}$], and object detection and segmentation [$\tikzcircle[black,fill=azure(colorwheel)]{2pt}$]. 
Supervised learning is common for data or visualization generation tasks as the ground-truth data or visualizations are usually provided for loss computation during training. It is also often used for prediction and object detection and segmentation tasks as the ground-truth results (e.g., future values, user-voted quality scores, segmentation masks) are given for network training. 
On the contrary, unsupervised learning is mostly applied for feature learning and extraction [$\tikzcircle[black,fill=electricpurple]{2pt}$]. AEs are often utilized for implicit feature learning from the input data in an unsupervised manner. 
In this case, these works exclusively belong to distributed learning, to be precise. 

Along the research task dimension, visualization generation tasks [$\tikzcircle[black,fill=gold(web)(golden)]{2pt}$] are exclusively supervised. 
Data generation [$\tikzcircle[black,fill=cherryblossompink]{2pt}$ $\tikzcircle[black,fill=brinkpink]{2pt}$ $\tikzcircle[black,fill=richcarmine]{2pt}$ $\tikzcircle[black,fill=carmine]{2pt}$], prediction [$\tikzcircle[black,fill=green-yellow]{2pt}$ $\tikzcircle[black,fill=green(pigment)]{2pt}$], and object detection and segmentation [$\tikzcircle[black,fill=azure(colorwheel)]{2pt}$] each occupy two learning types. 
Under the data generation category, two exceptions (i.e.,~\cite{Liu-JOV19,Lu-CGF21}) of compression and reconstruction [$\tikzcircle[black,fill=brinkpink]{2pt}$] fall into the category of distributed learning (a subtype of unsupervised learning). 
Under the prediction category, one exception (i.e.,~\cite{He-VI20}) of data-relevant prediction [$\tikzcircle[black,fill=green-yellow]{2pt}$] falls into distributed learning. 
Under the object detection and segmentation category, there are two exceptions (i.e.,\cite{He-JOV22-2,Nguyen-arXiv21}) of semi-supervised learning. 
Finally, feature learning and extraction tasks [$\tikzcircle[black,fill=electricpurple]{2pt}$] are most diverse across the learning types, covering all but weakly-supervised and disentangled learning. 

Only six examples fall into the categories of semi-supervised (i.e.,~\cite{He-JOV22-2,Nguyen-arXiv21,Zhu-CGA21}) and self-supervised (i.e.,~\cite{Raji-EGPGV17,Tkachev-TVCG,Chu-TOG17}) learning. 
Under self-supervised learning, all three works (i.e.,~\cite{Raji-EGPGV17,Tkachev-TVCG,Chu-TOG17}) are contrastive learning, not generative learning. 
Furthermore, two works (i.e.,~\cite{Han-VIS21,Nguyen-arXiv21}) employ pre-training (which aims to improve the network's generalization ability from the training datasets). Finally, two works (i.e.,~\cite{Cheng-TVCG19,Wang-VIS19}) use the pre-trained models directly to infer feature vectors from data. 
We find no existing DL4SciVis works in the disentangled learning, generative learning, and weakly-supervised learning types.  
This may be due to the challenges of defining and interpreting content and style for SciVis data, the high training and memory cost for the generative tasks, and the lack of weakly-supervised learning scenarios for SciVis, respectively.

\begin{table*}[t]
\caption{All surveyed papers and their neural network operations/connections and structures. The $^\triangleright$ sign indicates the work adds a discriminator. Specific network names other than their category names, if available, are listed after each paper. Color coding is for data generation [$\tikzcircle[black,fill=cherryblossompink]{2pt}$ $\tikzcircle[black,fill=brinkpink]{2pt}$ $\tikzcircle[black,fill=richcarmine]{2pt}$ $\tikzcircle[black,fill=carmine]{2pt}$], visualization generation [$\tikzcircle[black,fill=gold(web)(golden)]{2pt}$], prediction [$\tikzcircle[black,fill=green-yellow]{2pt}$ $\tikzcircle[black,fill=green(pigment)]{2pt}$], object detection and segmentation [$\tikzcircle[black,fill=azure(colorwheel)]{2pt}$], and feature learning and extraction [$\tikzcircle[black,fill=electricpurple]{2pt}$].}
\vspace{-0.05in}
\label{tab:network}
\centering
{\scriptsize
\begin{tabular}{c|c|c|c}
\hline
 operation/ & \multicolumn{3}{c}{structure} \\ \cline{2-4}
connection & encoder & decoder & encoder+decoder  \\ \hline
\multirow{7}{*}{CNN}  
& 
$\tikzcircle[black,fill=green(pigment)]{2pt}$\cite{Shi-TIST19}, 
$\tikzcircle[black,fill=green(pigment)]{2pt}$\cite{Yang-JOV19}  
& 
 $\tikzcircle[black,fill=cherryblossompink]{2pt}$\cite{An-CGA21}(U-Net, deformable CNN),
 $\tikzcircle[black,fill=cherryblossompink]{2pt}$\cite{Guo-PVIS20}
& 
$\tikzcircle[black,fill=azure(colorwheel)]{2pt}$\cite{Berenjkoub-VISSP20}(CNN vs.\ ResNet vs.\ U-Net), 
$\tikzcircle[black,fill=azure(colorwheel)]{2pt}$\cite{Borkiewicz-VISSP21}(U-Net),
$\tikzcircle[black,fill=richcarmine]{2pt}$\cite{Chu-TOG21}$^\triangleright$(GAN)
\\
& 
& 
 $\tikzcircle[black,fill=brinkpink]{2pt}$\cite{Han-CGA19}
 $\tikzcircle[black,fill=cherryblossompink]{2pt}$\cite{Han-TVCG}(GAN),
 $\tikzcircle[black,fill=cherryblossompink]{2pt}$\cite{Han-VIS21}$^\triangleright$(GAN)
& 
$\tikzcircle[black,fill=green(pigment)]{2pt}$\cite{Engel-VIS20},  
$\tikzcircle[black,fill=azure(colorwheel)]{2pt}$\cite{Ghahremani-TVCG}(nested encoder+decoder),
$\tikzcircle[black,fill=brinkpink]{2pt}$\cite{Gu-CGA21},
$\tikzcircle[black,fill=richcarmine]{2pt}$\cite{Gu-PVIS22},
$\tikzcircle[black,fill=cherryblossompink]{2pt}$\cite{Han-CG22}
\\
& 
& 
 $\tikzcircle[black,fill=cherryblossompink]{2pt}$\cite{Jakob-VIS20}(ESPCN),
 $\tikzcircle[black,fill=brinkpink]{2pt}$\cite{Liu-JOV19}$^\triangleright$(GAN), 
 $\tikzcircle[black,fill=cherryblossompink]{2pt}$\cite{Sahoo-EVISSP21}
& 
$\tikzcircle[black,fill=brinkpink]{2pt}$\cite{Han-VI22}$^\triangleright$(GAN),
$\tikzcircle[black,fill=richcarmine]{2pt}$\cite{Han-VIS20}$^\triangleright$(GAN),
$\tikzcircle[black,fill=gold(web)(golden)]{2pt}$\cite{Hong-PVIS19}$^\triangleright$(GAN)
\\
& 
& 
 $\tikzcircle[black,fill=gold(web)(golden)]{2pt}$\cite{Weiss-TVCG21}$^\triangleright$(FRVSR-Net),
 $\tikzcircle[black,fill=gold(web)(golden)]{2pt}$\cite{Weiss-TVCG}(EnhanceNet)
& 
$\tikzcircle[black,fill=azure(colorwheel)]{2pt}$\cite{Kashir-JOV21}(symmetric FCN),
$\tikzcircle[black,fill=richcarmine]{2pt}$\cite{Kim-CGF19-1}(AE),
$\tikzcircle[black,fill=electricpurple]{2pt}$\cite{Liu-arXiv19}(residual AE)
\\
& 
& 
 $\tikzcircle[black,fill=cherryblossompink]{2pt}$\cite{Werhahn-CGIT19}$^\triangleright$(multi-pass GAN) 
& 
$\tikzcircle[black,fill=azure(colorwheel)]{2pt}$\cite{Nguyen-arXiv21}(U-Net vs.\ U-Net+ResNet vs.\ DenseNet)
\\
& 
& 
 $\tikzcircle[black,fill=cherryblossompink]{2pt}$\cite{Wurster-arXiv21}$^\triangleright$(ESRGAN, WGAN)
& 
$\tikzcircle[black,fill=electricpurple]{2pt}$\cite{Porter-VISSP19}(AE),
$\tikzcircle[black,fill=azure(colorwheel)]{2pt}$\cite{Wang-VIS20}(multi-stream CNN)
\\
& 
& 
 $\tikzcircle[black,fill=cherryblossompink]{2pt}$\cite{Xie-TOG18}$^\triangleright$(GAN),  
 $\tikzcircle[black,fill=cherryblossompink]{2pt}$\cite{Zhou-CGI17} 
& 
$\tikzcircle[black,fill=gold(web)(golden)]{2pt}$\cite{Weiss-arXiv21-1}(U-Net, V-Net),
$\tikzcircle[black,fill=carmine]{2pt}$\cite{Wiewel-CGF19}, 
$\tikzcircle[black,fill=carmine]{2pt}$\cite{Wiewel-CGF20}
\\ \hline
MLP  
& 
$\tikzcircle[black,fill=green-yellow]{2pt}$\cite{Han-arXiv21},
$\tikzcircle[black,fill=electricpurple]{2pt}$\cite{Zhu-CGA21}(FCCNN) 
& 
---
& 
$\tikzcircle[black,fill=richcarmine]{2pt}$\cite{Kim-CGF19-1},
$\tikzcircle[black,fill=brinkpink]{2pt}$\cite{Lu-CGF21},
$\tikzcircle[black,fill=brinkpink]{2pt}$\cite{Weiss-arXiv21-2}
\\\hline
RNN  
& 
$\tikzcircle[black,fill=carmine]{2pt}$\cite{Wiewel-CGF20}(stacked LSTM) 
& 
$\tikzcircle[black,fill=carmine]{2pt}$\cite{Wiewel-CGF19}(LSTM) 
& 
---
\\\hline
GNN
& 
$\tikzcircle[black,fill=electricpurple]{2pt}$\cite{Han-CGF22}(GCN),
$\tikzcircle[black,fill=electricpurple]{2pt}$\cite{He-JOV22-1}(GCN) 
& 
$\tikzcircle[black,fill=brinkpink]{2pt}$\cite{Shi-PVIS22}(GCN)
& 
---
\\ \hline
\multirow{4}{*}{CNN+MLP} 
& 
$\tikzcircle[black,fill=electricpurple]{2pt}$\cite{Cheng-TVCG19},  
$\tikzcircle[black,fill=electricpurple]{2pt}$\cite{Chu-TOG17}(Siamese)
& 
$\tikzcircle[black,fill=gold(web)(golden)]{2pt}$\cite{He-VIS19}$^\triangleright$
& 
$\tikzcircle[black,fill=gold(web)(golden)]{2pt}$\cite{Berger-TVCG19}$^\triangleright$(GAN),  
$\tikzcircle[black,fill=electricpurple]{2pt}$\cite{Han-TVCG20}(AE)
\\
& 
$\tikzcircle[black,fill=azure(colorwheel)]{2pt}$\cite{Deng-JOV19},
$\tikzcircle[black,fill=green-yellow]{2pt}$\cite{He-VI20},  
$\tikzcircle[black,fill=green-yellow]{2pt}$\cite{Kim-CGF19-2}
& 
& 
$\tikzcircle[black,fill=electricpurple]{2pt}$\cite{Li-TVCG}(Geo-CNN)
\\
& 
$\tikzcircle[black,fill=electricpurple]{2pt}$\cite{Raji-EGPGV17}(Siamese), 
$\tikzcircle[black,fill=green-yellow]{2pt}$\cite{Tkachev-TVCG21}
& 
& 
\\
& 
$\tikzcircle[black,fill=electricpurple]{2pt}$\cite{Tkachev-TVCG}(Siamese),   
$\tikzcircle[black,fill=brinkpink]{2pt}$\cite{Wang-VIS19}   
& 
& 
\\ \hline
CNN+RNN
& 
---
& 
---
& 
$\tikzcircle[black,fill=cherryblossompink]{2pt}$\cite{Han-VIS19}$^\triangleright$(GAN+ConvLSTM)
\\ \hline
MLP+RNN
& 
$\tikzcircle[black,fill=green-yellow]{2pt}$\cite{Hong-PVIS18}(LSTM)
& 
---
& 
---
\\ \hline
GNN+MLP
& 
$\tikzcircle[black,fill=azure(colorwheel)]{2pt}$\cite{He-JOV22-2}(GCN) 
& 
---
& 
---
\\ \hline
\end{tabular}
}
\end{table*}

\begin{figure}[t]
\begin{center}
\includegraphics[width=1.0\linewidth]{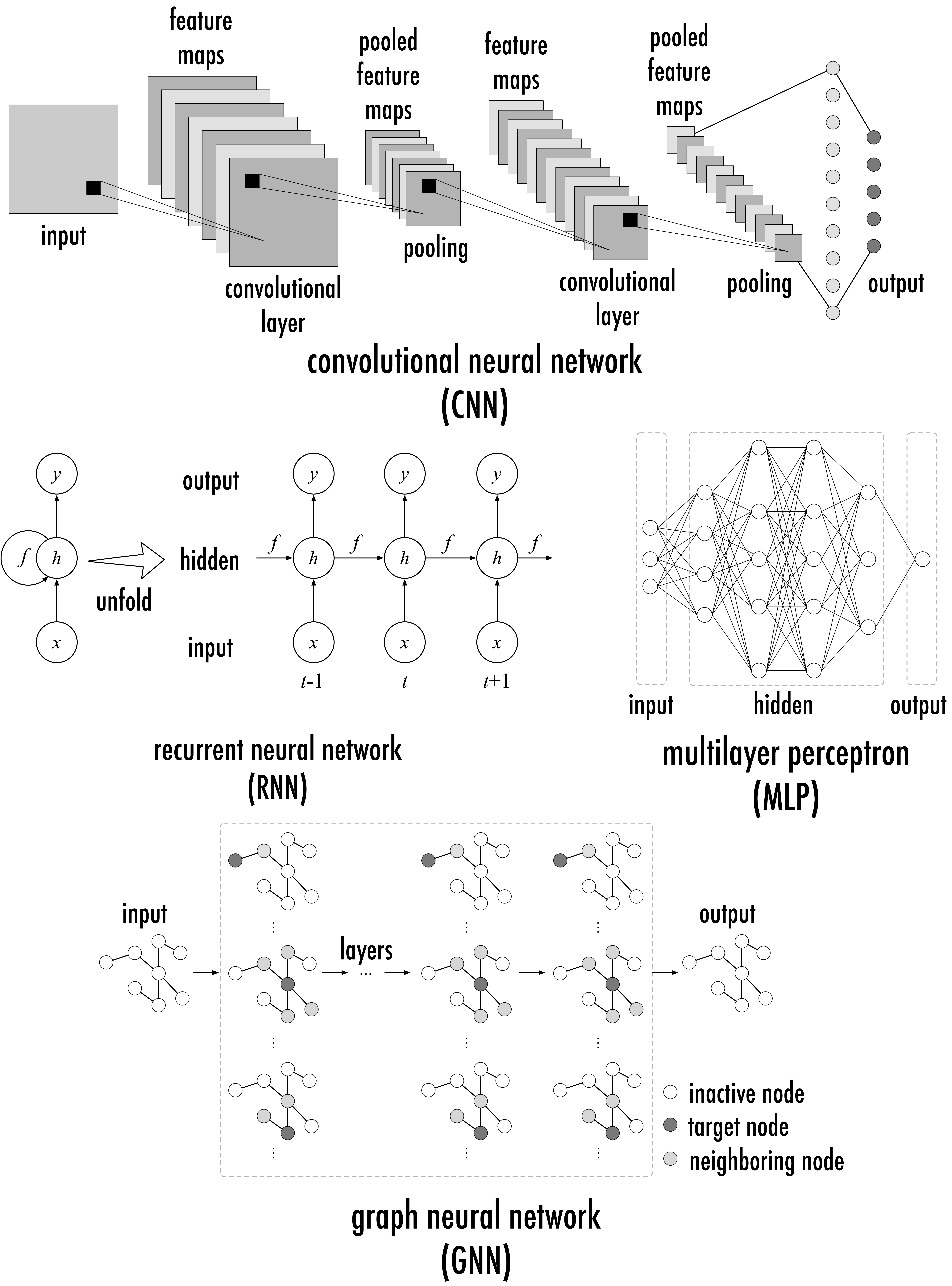}
\end{center}
\vspace{-0.1in}
\caption{Illustrative examples of CNN (redrawn from~\cite{Lecun-IEEE98}), RNN (adapted from~\cite{RNN-PIC}), MLP, and GNN network architectures.}
\label{fig:network}
\end{figure}

\vspace{-0.1in}
\subsection{Network Architectures} 
\label{subsec:na}
\vspace{-0.025in}

We categorize all surveyed papers based on their respective network operations and structures. 
From the network operation or connection perspective, there are four basic ones: {\em convolution operation}, {\em full connection}, {\em recurrence operation}, and {\em graph connection}, which lead to CNN, MLP, RNN, and GNN, respectively (refer to Figure~\ref{fig:network}). 
CNNs~\cite{Lecun-IEEE98} operate on grid-like data (e.g., images or volumes), where the convolution operation uses a convolution kernel or filter that slides along the input data to generate feature maps. 
The hidden layers of a CNN include several stages of convolutional and pooling layers, followed by one or several fully-connected layers. 
The {\em convolutional layer} detects local combinations of features from the previous layer. The {\em pooling layer} merges semantically similar features into one by computing a summary (e.g., maximum). The {\em fully-connected layer} learns global features from local ones by connecting neurons to all activations in the previous layer. 
RNNs~\cite{Rumelhart-TR85} operate on sequence data, where network connections between nodes form a directed graph along a temporal sequence. One of the most frequently used deep RNN architectures is LSTM~\cite{Hochreiter-NC97}, which was developed to address the vanishing gradient problem in traditional RNNs by adding forget gates in the units.  
MLPs~\cite{Rosenblatt-TR61} are networks composed of multiple layers of perceptrons. The input layer, one or multiple hidden layers, and the output layer are all fully connected (i.e., every node in one layer connects with a certain weight to each node in the following layer). 
Finally, GNNs~\cite{Gori-IJCNN05} naturally operate on graph-like data (e.g., surfaces). 
As a generalized version of CNN, GCN~\cite{Kipf-ICLR17} works on data with underlying non-regular structures. Each layer loops each node as a target node and applies a convolution operation on its neighboring nodes to learn node features. 

From the network structure perspective, there are three basic ones: {\em encoder}, {\em decoder}, and {\em encoder+decoder}. 
An encoder is a neural network that takes the data or rendering as input and outputs a reduced representation (e.g., feature vector) or estimated information (e.g., viewpoint) in a designated format. 
A decoder is also a neural network, usually the same as the encoder but in the opposite orientation. 
The decoder takes the coded message (i.e., reduced representation or estimated information) as input and outputs the data or infers the rendering result. 
An encoder+decoder includes both an encoder and a decoder. 
In the AE, the encoder is trained with the decoder.
The output feature vector from the encoder implicitly represents the information of the input data in the latent space. 
The decoder takes the feature vectors as input and produces the reconstructed data best matching the original data.
In other settings, it can complete a different task other than data reconstruction, such as super-resolution generation or data translation. 
The same technique has been used in various applications such as speech translation, generative models, etc.

Table~\ref{tab:network} shows our categorization of all surveyed papers. 
In addition, other or specific networks (e.g., GAN, deformable CNN) employed different from their categorical names are provided in the table. 
As an exception, only three works (i.e.,~\cite{Kim-CGF19-1,Wiewel-CGF19,Wiewel-CGF20}) are placed in two categories due to their use of separate networks. 

In terms of network operation or connection, we can see that CNN is the most popular category, followed by CNN+MLP. 
On the other hand, the least popular ones are GNN, CNN+RNN, MLP+RNN, and GNN+MLP. 
All four RNN-related works use only either LSTM~\cite{Hong-PVIS18,Wiewel-CGF19}, stacked LSTM~\cite{Wiewel-CGF20}, or ConvLSTM\cite{Han-VIS19}. The only four GNN-related works~\cite{He-JOV22-1,He-JOV22-2,Han-CGF22,Shi-PVIS22} use GCN.

In terms of network structure, encoder+decoder is most popular, and encoder and decoder are similarly popular. 
Among these three structures, encoder is mostly used for prediction [$\tikzcircle[black,fill=green-yellow]{2pt}$ $\tikzcircle[black,fill=green(pigment)]{2pt}$] and feature learning and extraction [$\tikzcircle[black,fill=electricpurple]{2pt}$] tasks, decoder is exclusively used for generation tasks (i.e., data generation [$\tikzcircle[black,fill=cherryblossompink]{2pt}$ $\tikzcircle[black,fill=brinkpink]{2pt}$ $\tikzcircle[black,fill=carmine]{2pt}$], visualization generation [$\tikzcircle[black,fill=gold(web)(golden)]{2pt}$]), while encoder+decoder serves all different categories of tasks. 
These are expected due to the nature of their respective functional roles of network structures: encoder performs compression, decode performs decompression, and as the combination of encoder and decoder, encoder+decoder is capable of accomplishing a variety of tasks. 

A close look at the categorization under CNN shows that encoder+decoder is the most diverse, covering all five research tasks, while decoder covers two tasks (i.e., data generation [$\tikzcircle[black,fill=cherryblossompink]{2pt}$ $\tikzcircle[black,fill=brinkpink]{2pt}$ $\tikzcircle[black,fill=carmine]{2pt}$], visualization generation [$\tikzcircle[black,fill=gold(web)(golden)]{2pt}$]), and encoder only covers the prediction task [$\tikzcircle[black,fill=green(pigment)]{2pt}$]. 
Many networks are either variants of CNNs or based on CNNs, including deformable CNN~\cite{Wang-CVPRW19}, 
DenseNet~\cite{Huang-CVPR17}, EnhanceNet~\cite{Sajjadi-ICCV17}, ESPCN~\cite{Shi-CVPR16}, FRVSR-Net~\cite{Sajjadi-CVPR18}, multi-stream CNN~\cite{Wang-VIS20}, ResNet~\cite{He-CVPR16}, symmetric FCN~\cite{Kashir-JOV21}, U-Net~\cite{Ronneberger-MICCAI15}, and V-Net~\cite{Milletari-3DV16}. 

For the categorization under CNN+MLP, encoder is the most diverse, covering four research tasks (i.e., data generation [$\tikzcircle[black,fill=brinkpink]{2pt}$], prediction [$\tikzcircle[black,fill=green-yellow]{2pt}$], object detection and segmentation [$\tikzcircle[black,fill=azure(colorwheel)]{2pt}$], feature learning and extraction [$\tikzcircle[black,fill=electricpurple]{2pt}$]). 
In comparison, encoder+decoder covers two tasks (i.e., 
visualization generation [$\tikzcircle[black,fill=gold(web)(golden)]{2pt}$], feature learning and extraction [$\tikzcircle[black,fill=electricpurple]{2pt}$]), and decoder only covers the visualization generation task [$\tikzcircle[black,fill=gold(web)(golden)]{2pt}$]. 

Explicitly set up to optimize for generative tasks, GAN covers both decoder and encoder+decoder categories, but not encoder. A GAN consists of two networks: a {\em generator} and a {\em discriminator}, which contest with each other in a zero-sum game. The generator maps from a latent space to a particular data distribution of interest. The discriminator discriminates between instances from the true data distribution and candidates produced by the generator. 
All works using a discriminator utilize GAN, except for one work~\cite{He-VI20} which employs a discriminative network instead of a GAN.

\begin{table*}[t]
\caption{All surveyed papers and their optimization targets and loss functions. Specific loss function names other than their category names, if available, are listed after each paper. In the ``others'' category, loss function names are listed before each paper. ``adversarial'' and ``compression'' are abbreviated as ``adv'' and ``comp'', respectively. Color coding is for data generation [$\tikzcircle[black,fill=cherryblossompink]{2pt}$ $\tikzcircle[black,fill=brinkpink]{2pt}$ $\tikzcircle[black,fill=richcarmine]{2pt}$ $\tikzcircle[black,fill=carmine]{2pt}$], visualization generation [$\tikzcircle[black,fill=gold(web)(golden)]{2pt}$], prediction [$\tikzcircle[black,fill=green-yellow]{2pt}$ $\tikzcircle[black,fill=green(pigment)]{2pt}$], object detection and segmentation [$\tikzcircle[black,fill=azure(colorwheel)]{2pt}$], and feature learning and extraction [$\tikzcircle[black,fill=electricpurple]{2pt}$].}
\vspace{-0.05in}
\label{tab:loss}
\centering
{\scriptsize
\begin{tabular}{c|c|c|c|c|c}
\hline
 & \multicolumn{5}{c}{loss} \\ \cline{2-6}
target & L1 (MAE) & L2 (MSE) & BCE & Wasserstein & others  \\ \hline

\multirow{7}{*}{data}  
& 
$\tikzcircle[black,fill=cherryblossompink]{2pt}$\cite{An-CGA21}(magnitude)
& 
$\tikzcircle[black,fill=richcarmine]{2pt}$\cite{Gu-PVIS22}(magnitude),
$\tikzcircle[black,fill=cherryblossompink]{2pt}$\cite{Guo-PVIS20}(magnitude)
& 
$\tikzcircle[black,fill=cherryblossompink]{2pt}$\cite{Han-VIS19}(adv)
& 
$\tikzcircle[black,fill=richcarmine]{2pt}$\cite{Chu-TOG21}(adv)
& 
CD$^2$ $\tikzcircle[black,fill=cherryblossompink]{2pt}$\cite{An-CGA21}(angle)
\\ & 
$\tikzcircle[black,fill=cherryblossompink]{2pt}$\cite{An-CGA21}(temporal),
$\tikzcircle[black,fill=richcarmine]{2pt}$\cite{Han-VIS20}(feature)
& 
$\tikzcircle[black,fill=cherryblossompink]{2pt}$\cite{Han-TVCG}(adv),
$\tikzcircle[black,fill=cherryblossompink]{2pt}$\cite{Han-TVCG}(content),
$\tikzcircle[black,fill=cherryblossompink]{2pt}$\cite{Han-TVCG}(feature)
& 
$\tikzcircle[black,fill=brinkpink]{2pt}$\cite{Han-VI22}(adv)
& 
$\tikzcircle[black,fill=cherryblossompink]{2pt}$\cite{Werhahn-CGIT19}(adv)
& 
CD$^2$ $\tikzcircle[black,fill=richcarmine]{2pt}$\cite{Gu-PVIS22}(angle)
\\ & 
$\tikzcircle[black,fill=gold(web)(golden)]{2pt}$\cite{Hong-PVIS19},
$\tikzcircle[black,fill=richcarmine]{2pt}$\cite{Kim-CGF19-1}(stream function)
& 
$\tikzcircle[black,fill=cherryblossompink]{2pt}$\cite{Han-VIS19}(feature),
$\tikzcircle[black,fill=cherryblossompink]{2pt}$\cite{Han-VIS19}(volumetric),
$\tikzcircle[black,fill=cherryblossompink]{2pt}$\cite{Han-CG22},
$\tikzcircle[black,fill=brinkpink]{2pt}$\cite{Han-VI22}
& 
$\tikzcircle[black,fill=gold(web)(golden)]{2pt}$\cite{Hong-PVIS19}(adv)
& 
$\tikzcircle[black,fill=cherryblossompink]{2pt}$\cite{Wurster-arXiv21}(adv)
& 
CD$^2$ $\tikzcircle[black,fill=cherryblossompink]{2pt}$\cite{Guo-PVIS20}(angle)
\\ & 
$\tikzcircle[black,fill=electricpurple]{2pt}$\cite{Liu-arXiv19},
$\tikzcircle[black,fill=brinkpink]{2pt}$\cite{Liu-JOV19},
$\tikzcircle[black,fill=brinkpink]{2pt}$\cite{Shi-PVIS22}
& 
$\tikzcircle[black,fill=cherryblossompink]{2pt}$\cite{Han-VIS21}(adv),
$\tikzcircle[black,fill=cherryblossompink]{2pt}$\cite{Han-VIS21}(cycle),
$\tikzcircle[black,fill=cherryblossompink]{2pt}$\cite{Han-VIS21}(volumetric)
& 
$\tikzcircle[black,fill=brinkpink]{2pt}$\cite{Liu-JOV19}(adv)
& 
& 
\\ & 
$\tikzcircle[black,fill=cherryblossompink]{2pt}$\cite{Werhahn-CGIT19},
$\tikzcircle[black,fill=carmine]{2pt}$\cite{Wiewel-CGF19}
& 
$\tikzcircle[black,fill=richcarmine]{2pt}$\cite{Han-VIS20}(adv),
$\tikzcircle[black,fill=richcarmine]{2pt}$\cite{Han-VIS20}(volumetric)
& 
$\tikzcircle[black,fill=cherryblossompink]{2pt}$\cite{Xie-TOG18}(adv)
& 
& 
\\ & 
$\tikzcircle[black,fill=carmine]{2pt}$\cite{Wiewel-CGF20}(AE),
$\tikzcircle[black,fill=carmine]{2pt}$\cite{Wiewel-CGF20}(split)
& 
$\tikzcircle[black,fill=brinkpink]{2pt}$\cite{Lu-CGF21}(reconstruction),
$\tikzcircle[black,fill=electricpurple]{2pt}$\cite{Porter-VISSP19}
& 
& 
& 
\\ & 
$\tikzcircle[black,fill=cherryblossompink]{2pt}$\cite{Wurster-arXiv21}(reconstruction),
$\tikzcircle[black,fill=cherryblossompink]{2pt}$\cite{Xie-TOG18}
& 
$\tikzcircle[black,fill=cherryblossompink]{2pt}$\cite{Sahoo-EVISSP21}(vector),
$\tikzcircle[black,fill=green-yellow]{2pt}$\cite{Tkachev-TVCG21},
$\tikzcircle[black,fill=carmine]{2pt}$\cite{Wiewel-CGF19},
$\tikzcircle[black,fill=cherryblossompink]{2pt}$\cite{Zhou-CGI17}
& 
& 
& 
\\
\hline
\multirow{2}{*}{image}  
& 
$\tikzcircle[black,fill=gold(web)(golden)]{2pt}$\cite{Berger-TVCG19},
$\tikzcircle[black,fill=brinkpink]{2pt}$\cite{Weiss-arXiv21-2}
& 
$\tikzcircle[black,fill=gold(web)(golden)]{2pt}$\cite{Weiss-arXiv21-1}
& 
$\tikzcircle[black,fill=gold(web)(golden)]{2pt}$\cite{Berger-TVCG19}(adv)
& 
---
& 
SSIM $\tikzcircle[black,fill=gold(web)(golden)]{2pt}$\cite{Weiss-arXiv21-1}
\\ 
& 
& 
& 
$\tikzcircle[black,fill=gold(web)(golden)]{2pt}$\cite{He-VIS19}(adv)
& 
& 
\\ \hline
\multirow{9}{*}{feature}  
& 
$\tikzcircle[black,fill=cherryblossompink]{2pt}$\cite{An-CGA21}(Jacobian)
& 
$\tikzcircle[black,fill=gold(web)(golden)]{2pt}$\cite{Berger-TVCG19}(AE),
$\tikzcircle[black,fill=electricpurple]{2pt}$\cite{Chu-TOG17}(hinge),
$\tikzcircle[black,fill=green(pigment)]{2pt}$\cite{Engel-VIS20}
& 
$\tikzcircle[black,fill=gold(web)(golden)]{2pt}$\cite{Weiss-TVCG}
& 
---
& 
SSIM $\tikzcircle[black,fill=green(pigment)]{2pt}$\cite{Engel-VIS20}(AO)
\\ 
& 
$\tikzcircle[black,fill=richcarmine]{2pt}$\cite{Chu-TOG21}(modified)
& 
$\tikzcircle[black,fill=brinkpink]{2pt}$\cite{Gu-CGA21}(streamline),
$\tikzcircle[black,fill=richcarmine]{2pt}$\cite{Gu-PVIS22}(Jacobian)
& 
& 
& 
contrastive $\tikzcircle[black,fill=electricpurple]{2pt}$\cite{He-JOV22-1}(local)
\\ 
& 
$\tikzcircle[black,fill=richcarmine]{2pt}$\cite{Chu-TOG21}(regularized),
$\tikzcircle[black,fill=green-yellow]{2pt}$\cite{Han-arXiv21}
& 
$\tikzcircle[black,fill=brinkpink]{2pt}$\cite{Han-CGA19}(streamline),
$\tikzcircle[black,fill=electricpurple]{2pt}$\cite{Han-CGF22}
& 
& 
& 
MI $\tikzcircle[black,fill=electricpurple]{2pt}$\cite{He-JOV22-1}(global)
\\ 
& 
$\tikzcircle[black,fill=richcarmine]{2pt}$\cite{Kim-CGF19-1}(stream function)
& 
$\tikzcircle[black,fill=gold(web)(golden)]{2pt}$\cite{He-VIS19}(reconstruction),
$\tikzcircle[black,fill=cherryblossompink]{2pt}$\cite{Jakob-VIS20},
$\tikzcircle[black,fill=richcarmine]{2pt}$\cite{Kim-CGF19-1},
$\tikzcircle[black,fill=green-yellow]{2pt}$\cite{Kim-CGF19-2}
& 
& 
& 
CD$^1$ $\tikzcircle[black,fill=brinkpink]{2pt}$\cite{Wang-VIS19}(deformation)
\\ 
& 
$\tikzcircle[black,fill=richcarmine]{2pt}$\cite{Kim-CGF19-1}(velocity gradient)
& 
$\tikzcircle[black,fill=electricpurple]{2pt}$\cite{Li-TVCG}(attribute),
$\tikzcircle[black,fill=brinkpink]{2pt}$\cite{Lu-CGF21}(scalar gradient),
$\tikzcircle[black,fill=electricpurple]{2pt}$\cite{Raji-EGPGV17}
& 
& 
& 
\\ 
& 
$\tikzcircle[black,fill=electricpurple]{2pt}$\cite{Tkachev-TVCG},
$\tikzcircle[black,fill=gold(web)(golden)]{2pt}$\cite{Weiss-TVCG21}(AO)
& 
$\tikzcircle[black,fill=cherryblossompink]{2pt}$\cite{Sahoo-EVISSP21}(streamline),
$\tikzcircle[black,fill=brinkpink]{2pt}$\cite{Wang-VIS19}(regularization)
& 
& 
& 
\\ 
& 
$\tikzcircle[black,fill=gold(web)(golden)]{2pt}$\cite{Weiss-TVCG21}(depth),
$\tikzcircle[black,fill=gold(web)(golden)]{2pt}$\cite{Weiss-TVCG21}(mask)
& 
$\tikzcircle[black,fill=brinkpink]{2pt}$\cite{Wang-VIS19}(translation),
$\tikzcircle[black,fill=gold(web)(golden)]{2pt}$\cite{Weiss-TVCG21}(color-temporal)
& 
& 
& 
\\ 
& 
$\tikzcircle[black,fill=gold(web)(golden)]{2pt}$\cite{Weiss-TVCG21}(normal)
& 
$\tikzcircle[black,fill=gold(web)(golden)]{2pt}$\cite{Weiss-TVCG}(bounds),
$\tikzcircle[black,fill=gold(web)(golden)]{2pt}$\cite{Weiss-TVCG}(prior)
& 
& 
& 
\\ 
& 
$\tikzcircle[black,fill=gold(web)(golden)]{2pt}$\cite{Weiss-TVCG},
$\tikzcircle[black,fill=carmine]{2pt}$\cite{Wiewel-CGF20}(AE)
& 
$\tikzcircle[black,fill=carmine]{2pt}$\cite{Wiewel-CGF20}(AE),
$\tikzcircle[black,fill=carmine]{2pt}$\cite{Wiewel-CGF20}(supervised),
$\tikzcircle[black,fill=cherryblossompink]{2pt}$\cite{Xie-TOG18}
& 
& 
& 
\\ \hline
\multirow{8}{*}{probability}  
& 
---
& 
$\tikzcircle[black,fill=electricpurple]{2pt}$\cite{Han-TVCG20},
$\tikzcircle[black,fill=azure(colorwheel)]{2pt}$\cite{Nguyen-arXiv21},
$\tikzcircle[black,fill=green(pigment)]{2pt}$\cite{Yang-JOV19}
& 
$\tikzcircle[black,fill=azure(colorwheel)]{2pt}$\cite{Berenjkoub-VISSP20},
$\tikzcircle[black,fill=azure(colorwheel)]{2pt}$\cite{Borkiewicz-VISSP21}
& 
$\tikzcircle[black,fill=green-yellow]{2pt}$\cite{He-VI20}
& 
CE $\tikzcircle[black,fill=electricpurple]{2pt}$\cite{Cheng-TVCG19},
Dice $\tikzcircle[black,fill=electricpurple]{2pt}$\cite{Han-TVCG20}
\\
& 
& 
& 
$\tikzcircle[black,fill=azure(colorwheel)]{2pt}$\cite{Deng-JOV19},
$\tikzcircle[black,fill=azure(colorwheel)]{2pt}$\cite{Ghahremani-TVCG}
& 
& 
CE $\tikzcircle[black,fill=azure(colorwheel)]{2pt}$\cite{He-JOV22-2}
\\
& 
& 
& 
$\tikzcircle[black,fill=electricpurple]{2pt}$\cite{Han-TVCG20},
$\tikzcircle[black,fill=azure(colorwheel)]{2pt}$\cite{Kashir-JOV21}
& 
& 
log-likehood $\tikzcircle[black,fill=green-yellow]{2pt}$\cite{Hong-PVIS18}
\\
& 
& 
& 
$\tikzcircle[black,fill=azure(colorwheel)]{2pt}$\cite{Nguyen-arXiv21}
& 
& 
AWL $\tikzcircle[black,fill=azure(colorwheel)]{2pt}$\cite{Nguyen-arXiv21}
\\
& 
& 
& 
$\tikzcircle[black,fill=electricpurple]{2pt}$\cite{Tkachev-TVCG}
& 
& 
CE $\tikzcircle[black,fill=green(pigment)]{2pt}$\cite{Shi-TIST19}(GSAL)
\\
& 
& 
& 
& 
& 
Dice $\tikzcircle[black,fill=azure(colorwheel)]{2pt}$\cite{Wang-VIS20}(MIP)
\\
& 
& 
& 
& 
& 
Dice $\tikzcircle[black,fill=azure(colorwheel)]{2pt}$\cite{Wang-VIS20}(voxel)
\\
& 
& 
& 
& 
& 
CE $\tikzcircle[black,fill=electricpurple]{2pt}$\cite{Zhu-CGA21}
\\ \hline
\multirow{2}{*}{parameter}  
& 
---
& 
$\tikzcircle[black,fill=green-yellow]{2pt}$\cite{He-VI20}(gradient penalty),
$\tikzcircle[black,fill=electricpurple]{2pt}$\cite{Tkachev-TVCG}
& 
---
& 
---
& 
---
\\
& 
& 
$\tikzcircle[black,fill=cherryblossompink]{2pt}$\cite{Werhahn-CGIT19}(gradient penalty)
& 
& 
& 
\\
\hline
\end{tabular}
}
\end{table*}

\vspace{-0.1in}
\subsection{Loss Functions} 
\vspace{-0.025in}

An essential aspect of training a neural network lies in the design of the objective functions. 
Usually, the training goal is to minimize the objective functions over multiple iterations or epochs until the network converges. In this scenario, the objective function is often called the loss function (a.k.a. cost function or error function). 

We categorize the optimization targets into five levels: {\em data}, {\em image}, {\em feature}, {\em probability}, and {\em parameter}. 
Data-level targets minimize the differences between model-inferred and ground-truth data (e.g., scalar or vector fields). The corresponding loss functions typically operate on individual voxels (3D data) or pixels (2D data) to accumulate the errors or compare the errors between two probability distributions (one from the mode-inferred data and another from the ground-truth data).
Image-level targets minimize the differences between the synthesized and ground-truth rendering results (e.g., volume rendering, isosurface rendering). Like data-level targets, their corresponding loss functions loop through pixels or compare the probability distributions or statistical quantities (e.g., mean, standard deviation). 
Feature-level targets look to minimize the differences between features, properties, or attributes (e.g., streamlines, gradients, AO) derived from the inferred and ground-truth data.
Probability-level targets aim to minimize the differences of predicted probabilities (e.g., segmentation masks) between the inferred and ground-truth results, typically in object detection and segmentation tasks. The loss functions often take the form of CE (multi-class classification or segmentation) or BCE (binary classification or segmentation), based on the framework of maximum likelihood.
Finally, parameter-level targets directly minimize the neural network parameters during training. The corresponding loss functions are generally related to regularization terms or gradient penalties. 

As shown in Table~\ref{tab:loss}, we group all surveyed papers based on their optimization targets (organized in primary rows). For loss functions, we single out four categories: L1 (MAE), L2 (MSE), BCE, and Wasserstein, and leave all remaining ones under the category of ``others''. 

L1 (MAE) and L2 (MSE) are the most popular loss functions. L1 loss minimizes the {\em absolute} differences between the inferred values and the ground-truth values, while L2 loss minimizes the {\em squared} differences between them. Because the difference between an incorrectly predicted value and the ground-truth value could be fairly large, squaring it would significantly amplify the difference. Therefore, compared with L2 loss, L1 loss is more stable and less susceptible to outliers. 

For training a generative model using GANs, standard adversarial loss functions are based on L2, BCE, or Wasserstein. 
In the binary classification or segmentation setting, BCE compares the predicted probabilities with the actual binary class output and penalizes the probabilities if the distances from the expected ones are large. 
Unlike divergence-based loss functions (e.g., Kullback-Leibler divergence~\cite{Kullback-AMS51}, Jensen-Shannon divergence~\cite{Lin-TIT51}), Wasserstein loss~\cite{Frogner-NIPS15} considers optimal transport by utilizing EMD as a natural distance for
probability distributions over metric spaces. As an alternative to traditional GAN training, WGAN~\cite{Arjovsky-ICML17} can improve the stability of the network's optimization process. 

Across the primary rows in Table~\ref{tab:loss}, we can see that data- and feature-level optimization targets are the most widely used ones, followed by probability-level targets. Finally, image- and parameter-level optimization targets are the least employed ones. 
Across the columns, we see that L2 (MSE) and L1 (MAE) losses are the most popular ones. 
Between BCE and Wasserstein, BCE is more frequently used due to its simplicity and easy implementation. 
In the column of ``others'', different loss functions other than L1, L2, BCE, and Wasserstein include CE, Dice, CD$^2$, SSIM, etc. 

Note that more than half of the surveyed papers employ more than one loss term. The goal is to consider different aspects to improve the overall inference quality. In addition, many papers coin specific loss names (e.g., adversarial, content, cycle, feature, reconstruction, temporal, and volumetric losses) reflecting their respective contexts. But in essence, the underlying loss function is mostly L1 or L2 loss.

Along the research task dimension, data generation tasks [$\tikzcircle[black,fill=cherryblossompink]{2pt}$ $\tikzcircle[black,fill=brinkpink]{2pt}$ $\tikzcircle[black,fill=richcarmine]{2pt}$ $\tikzcircle[black,fill=carmine]{2pt}$] employ losses across all levels except for probability-level targets. Visualization generation tasks [$\tikzcircle[black,fill=gold(web)(golden)]{2pt}$] use losses across data-, image-, and feature-level targets. Prediction tasks [$\tikzcircle[black,fill=green-yellow]{2pt}$ $\tikzcircle[black,fill=green(pigment)]{2pt}$] and feature learning and extraction tasks [$\tikzcircle[black,fill=electricpurple]{2pt}$] utilize losses across all levels except for image-level targets. Finally, object detection and segmentation tasks [$\tikzcircle[black,fill=azure(colorwheel)]{2pt}$] only use the probability-level optimization target. 

\begin{table}[t]
\caption{Categorization of the surveyed papers into the six levels of evaluation metrics. Metrics in each category are arranged based on popularity, followed by alphabetical order. Color coding is for data generation [$\tikzcircle[black,fill=cherryblossompink]{2pt}$ $\tikzcircle[black,fill=brinkpink]{2pt}$ $\tikzcircle[black,fill=richcarmine]{2pt}$ $\tikzcircle[black,fill=carmine]{2pt}$], visualization generation [$\tikzcircle[black,fill=gold(web)(golden)]{2pt}$], prediction [$\tikzcircle[black,fill=green-yellow]{2pt}$ $\tikzcircle[black,fill=green(pigment)]{2pt}$], object detection and segmentation [$\tikzcircle[black,fill=azure(colorwheel)]{2pt}$], and feature learning and extraction [$\tikzcircle[black,fill=electricpurple]{2pt}$].}
\vspace{-0.05in}
\label{tab:metrics}
\centering
{\scriptsize
\begin{tabular}{l|p{6.5cm}}
\hline
 level & metric: papers \\ \hline
 &
 PSNR: 
 $\tikzcircle[black,fill=cherryblossompink]{2pt}$\cite{An-CGA21}, 
 $\tikzcircle[black,fill=brinkpink]{2pt}$\cite{Gu-CGA21},
 $\tikzcircle[black,fill=richcarmine]{2pt}$\cite{Gu-PVIS22},
 $\tikzcircle[black,fill=cherryblossompink]{2pt}$\cite{Guo-PVIS20},
 $\tikzcircle[black,fill=electricpurple]{2pt}$\cite{Han-TVCG20}, 
 $\tikzcircle[black,fill=brinkpink]{2pt}$\cite{Han-CGA19}, 
 $\tikzcircle[black,fill=cherryblossompink]{2pt}$\cite{Han-TVCG}\\
 &
 \hspace{7mm} $\tikzcircle[black,fill=cherryblossompink]{2pt}$\cite{Han-VIS19},
 $\tikzcircle[black,fill=cherryblossompink]{2pt}$\cite{Han-CG22},
 $\tikzcircle[black,fill=brinkpink]{2pt}$\cite{Han-VI22},
 $\tikzcircle[black,fill=cherryblossompink]{2pt}$\cite{Han-VIS21},
 $\tikzcircle[black,fill=richcarmine]{2pt}$\cite{Han-VIS20}, 
 $\tikzcircle[black,fill=brinkpink]{2pt}$\cite{Lu-CGF21},
 $\tikzcircle[black,fill=electricpurple]{2pt}$\cite{Porter-VISSP19}\\
 &
  \hspace{7mm} 
 $\tikzcircle[black,fill=cherryblossompink]{2pt}$\cite{Sahoo-EVISSP21},
 $\tikzcircle[black,fill=brinkpink]{2pt}$\cite{Shi-PVIS22},
 $\tikzcircle[black,fill=carmine]{2pt}$\cite{Wiewel-CGF19},
 $\tikzcircle[black,fill=carmine]{2pt}$\cite{Wiewel-CGF20},
 $\tikzcircle[black,fill=cherryblossompink]{2pt}$\cite{Wurster-arXiv21},
 $\tikzcircle[black,fill=cherryblossompink]{2pt}$\cite{Zhou-CGI17}\\
& CR: 
$\tikzcircle[black,fill=cherryblossompink]{2pt}$\cite{An-CGA21}, 
$\tikzcircle[black,fill=richcarmine]{2pt}$\cite{Gu-PVIS22},
$\tikzcircle[black,fill=brinkpink]{2pt}$\cite{Han-CGA19}, 
$\tikzcircle[black,fill=cherryblossompink]{2pt}$\cite{Han-CG22},
$\tikzcircle[black,fill=richcarmine]{2pt}$\cite{Kim-CGF19-1}, 
$\tikzcircle[black,fill=brinkpink]{2pt}$\cite{Liu-JOV19}, 
$\tikzcircle[black,fill=brinkpink]{2pt}$\cite{Lu-CGF21}\\
data &
\hspace{4mm}
$\tikzcircle[black,fill=brinkpink]{2pt}$\cite{Weiss-arXiv21-2},
$\tikzcircle[black,fill=cherryblossompink]{2pt}$\cite{Wurster-arXiv21}\\
  &
 AAD: 
 $\tikzcircle[black,fill=brinkpink]{2pt}$\cite{Gu-CGA21}, 
 $\tikzcircle[black,fill=cherryblossompink]{2pt}$\cite{Guo-PVIS20},
 $\tikzcircle[black,fill=electricpurple]{2pt}$\cite{Han-TVCG20},
 $\tikzcircle[black,fill=cherryblossompink]{2pt}$\cite{Han-CG22}\\
&
 RMSE: 
 $\tikzcircle[black,fill=cherryblossompink]{2pt}$\cite{Guo-PVIS20},
 $\tikzcircle[black,fill=brinkpink]{2pt}$\cite{Liu-JOV19}, 
 $\tikzcircle[black,fill=electricpurple]{2pt}$\cite{Porter-VISSP19}
  SSIM: 
 $\tikzcircle[black,fill=electricpurple]{2pt}$\cite{Tkachev-TVCG}, 
 $\tikzcircle[black,fill=cherryblossompink]{2pt}$\cite{Wurster-arXiv21}, 
 $\tikzcircle[black,fill=cherryblossompink]{2pt}$\cite{Zhou-CGI17}\\
 &
 MAE: 
 $\tikzcircle[black,fill=richcarmine]{2pt}$\cite{Chu-TOG21}, 
 $\tikzcircle[black,fill=richcarmine]{2pt}$\cite{Kim-CGF19-1}
 MSE: 
 $\tikzcircle[black,fill=green-yellow]{2pt}$\cite{Tkachev-TVCG21}, 
 $\tikzcircle[black,fill=electricpurple]{2pt}$\cite{Tkachev-TVCG}
RAE: 
 $\tikzcircle[black,fill=richcarmine]{2pt}$\cite{Gu-PVIS22},
 $\tikzcircle[black,fill=cherryblossompink]{2pt}$\cite{Han-CG22}\\
 &
 CD$^2$: 
 $\tikzcircle[black,fill=electricpurple]{2pt}$\cite{He-JOV22-1}
 EMD: 
 $\tikzcircle[black,fill=electricpurple]{2pt}$\cite{Tkachev-TVCG}
 LSiM: 
 $\tikzcircle[black,fill=richcarmine]{2pt}$\cite{Chu-TOG21}
 VGG metric: 
 $\tikzcircle[black,fill=electricpurple]{2pt}$\cite{Tkachev-TVCG}
 \\ \hline
 &
 SSIM: 
 $\tikzcircle[black,fill=azure(colorwheel)]{2pt}$\cite{Ghahremani-TVCG}, 
 $\tikzcircle[black,fill=cherryblossompink]{2pt}$\cite{Han-TVCG}, 
 $\tikzcircle[black,fill=cherryblossompink]{2pt}$\cite{Han-VIS19}, 
 $\tikzcircle[black,fill=cherryblossompink]{2pt}$\cite{Han-VIS21}, 
 $\tikzcircle[black,fill=richcarmine]{2pt}$\cite{Han-VIS20},
 $\tikzcircle[black,fill=gold(web)(golden)]{2pt}$\cite{He-VIS19},
 $\tikzcircle[black,fill=brinkpink]{2pt}$\cite{Shi-PVIS22}\\
 &
 \hspace{6.25mm} $\tikzcircle[black,fill=gold(web)(golden)]{2pt}$\cite{Weiss-arXiv21-1},
 $\tikzcircle[black,fill=gold(web)(golden)]{2pt}$\cite{Weiss-TVCG21},
 $\tikzcircle[black,fill=brinkpink]{2pt}$\cite{Weiss-arXiv21-2},
 $\tikzcircle[black,fill=gold(web)(golden)]{2pt}$\cite{Weiss-TVCG}\\
  image &
 LPIPS: 
 $\tikzcircle[black,fill=cherryblossompink]{2pt}$\cite{Han-CG22},
 $\tikzcircle[black,fill=gold(web)(golden)]{2pt}$\cite{Weiss-arXiv21-1}, 
 $\tikzcircle[black,fill=brinkpink]{2pt}$\cite{Weiss-arXiv21-2}, 
 $\tikzcircle[black,fill=gold(web)(golden)]{2pt}$\cite{Weiss-TVCG}\\
 & 
 PSNR: 
 $\tikzcircle[black,fill=gold(web)(golden)]{2pt}$\cite{He-VIS19}, 
 $\tikzcircle[black,fill=gold(web)(golden)]{2pt}$\cite{Hong-PVIS19}, 
 $\tikzcircle[black,fill=gold(web)(golden)]{2pt}$\cite{Weiss-TVCG21}, 
 $\tikzcircle[black,fill=gold(web)(golden)]{2pt}$\cite{Weiss-TVCG}\\
 &
 EMD: 
 $\tikzcircle[black,fill=gold(web)(golden)]{2pt}$\cite{Berger-TVCG19}, 
 $\tikzcircle[black,fill=gold(web)(golden)]{2pt}$\cite{He-VIS19},
 $\tikzcircle[black,fill=brinkpink]{2pt}$\cite{Shi-PVIS22}
 FID: 
 $\tikzcircle[black,fill=gold(web)(golden)]{2pt}$\cite{He-VIS19}, 
 $\tikzcircle[black,fill=gold(web)(golden)]{2pt}$\cite{Weiss-arXiv21-1}\\
 &
 MAE: 
 $\tikzcircle[black,fill=azure(colorwheel)]{2pt}$\cite{Ghahremani-TVCG}
 REC: 
 $\tikzcircle[black,fill=gold(web)(golden)]{2pt}$\cite{Weiss-TVCG21}
 RMSE: 
 $\tikzcircle[black,fill=gold(web)(golden)]{2pt}$\cite{Berger-TVCG19}
 \\ \hline
 &
IS: 
 $\tikzcircle[black,fill=cherryblossompink]{2pt}$\cite{Han-TVCG},
 $\tikzcircle[black,fill=cherryblossompink]{2pt}$\cite{Han-VIS19}, 
 $\tikzcircle[black,fill=brinkpink]{2pt}$\cite{Han-VI22},
 $\tikzcircle[black,fill=cherryblossompink]{2pt}$\cite{Han-VIS21}, 
 $\tikzcircle[black,fill=richcarmine]{2pt}$\cite{Han-VIS20}\\
 &
 MSE: 
 $\tikzcircle[black,fill=azure(colorwheel)]{2pt}$\cite{Berenjkoub-VISSP20}, 
 $\tikzcircle[black,fill=electricpurple]{2pt}$\cite{Cheng-TVCG19}, 
 $\tikzcircle[black,fill=green(pigment)]{2pt}$\cite{Engel-VIS20},
 $\tikzcircle[black,fill=cherryblossompink]{2pt}$\cite{Jakob-VIS20}, 
 $\tikzcircle[black,fill=electricpurple]{2pt}$\cite{Zhu-CGA21}\\  
  &
 MCPD: 
 $\tikzcircle[black,fill=cherryblossompink]{2pt}$\cite{An-CGA21},
 $\tikzcircle[black,fill=brinkpink]{2pt}$\cite{Gu-CGA21},
 $\tikzcircle[black,fill=richcarmine]{2pt}$\cite{Gu-PVIS22},
 $\tikzcircle[black,fill=cherryblossompink]{2pt}$\cite{Han-CG22}\\
 &
 HD: 
 $\tikzcircle[black,fill=brinkpink]{2pt}$\cite{Wang-VIS19}, 
 $\tikzcircle[black,fill=carmine]{2pt}$\cite{Wiewel-CGF19}
 IOU:
 $\tikzcircle[black,fill=brinkpink]{2pt}$\cite{Shi-PVIS22},
 $\tikzcircle[black,fill=brinkpink]{2pt}$\cite{Wang-VIS19}\\
 feature &
 accuracy: 
 $\tikzcircle[black,fill=azure(colorwheel)]{2pt}$\cite{Ghahremani-TVCG}
 ACPPE: 
 $\tikzcircle[black,fill=cherryblossompink]{2pt}$\cite{An-CGA21}
 AEDR:
 $\tikzcircle[black,fill=green-yellow]{2pt}$\cite{Han-arXiv21}\\
 &
 ALP: 
 $\tikzcircle[black,fill=cherryblossompink]{2pt}$\cite{Sahoo-EVISSP21}
 association score: 
 $\tikzcircle[black,fill=electricpurple]{2pt}$\cite{He-JOV22-1}\\
 &
 CD$^1$: 
 $\tikzcircle[black,fill=brinkpink]{2pt}$\cite{Wang-VIS19}
 EMD: 
 $\tikzcircle[black,fill=brinkpink]{2pt}$\cite{Wang-VIS19}
 F-score:
 $\tikzcircle[black,fill=brinkpink]{2pt}$\cite{Wang-VIS19}\\
 &
 feature deviation:
 $\tikzcircle[black,fill=electricpurple]{2pt}$\cite{Li-TVCG}
 finger count: 
 $\tikzcircle[black,fill=electricpurple]{2pt}$\cite{Li-TVCG}\\
 &
 global error:
 $\tikzcircle[black,fill=green-yellow]{2pt}$\cite{Han-arXiv21}
 local error:
 $\tikzcircle[black,fill=green-yellow]{2pt}$\cite{Han-arXiv21}\\
 &
 ME: 
 $\tikzcircle[black,fill=electricpurple]{2pt}$\cite{Cheng-TVCG19}
 PSNR: 
 $\tikzcircle[black,fill=cherryblossompink]{2pt}$\cite{Jakob-VIS20}
 SC: 
 $\tikzcircle[black,fill=electricpurple]{2pt}$\cite{He-JOV22-1}
 SSIM:
  $\tikzcircle[black,fill=green(pigment)]{2pt}$\cite{Engel-VIS20}\\
 \hline
 &
 F-score: 
  $\tikzcircle[black,fill=azure(colorwheel)]{2pt}$\cite{Berenjkoub-VISSP20}, 
  $\tikzcircle[black,fill=azure(colorwheel)]{2pt}$\cite{Ghahremani-TVCG},
  $\tikzcircle[black,fill=electricpurple]{2pt}$\cite{Han-TVCG20},
  $\tikzcircle[black,fill=azure(colorwheel)]{2pt}$\cite{He-JOV22-2},
  $\tikzcircle[black,fill=azure(colorwheel)]{2pt}$\cite{Kashir-JOV21},
  $\tikzcircle[black,fill=azure(colorwheel)]{2pt}$\cite{Nguyen-arXiv21},
  $\tikzcircle[black,fill=azure(colorwheel)]{2pt}$\cite{Wang-VIS20}\\
 &
 PPV:
 $\tikzcircle[black,fill=azure(colorwheel)]{2pt}$\cite{Deng-JOV19}, 
 $\tikzcircle[black,fill=azure(colorwheel)]{2pt}$\cite{Ghahremani-TVCG},
 $\tikzcircle[black,fill=azure(colorwheel)]{2pt}$\cite{He-JOV22-2},
 $\tikzcircle[black,fill=azure(colorwheel)]{2pt}$\cite{Kashir-JOV21},
 $\tikzcircle[black,fill=azure(colorwheel)]{2pt}$\cite{Wang-VIS20}\\  
 probability &
  TPR:
 $\tikzcircle[black,fill=azure(colorwheel)]{2pt}$\cite{Deng-JOV19},
 $\tikzcircle[black,fill=azure(colorwheel)]{2pt}$\cite{Ghahremani-TVCG},
 $\tikzcircle[black,fill=azure(colorwheel)]{2pt}$\cite{He-JOV22-2}
  IOU: 
 $\tikzcircle[black,fill=azure(colorwheel)]{2pt}$\cite{Borkiewicz-VISSP21}, 
 $\tikzcircle[black,fill=azure(colorwheel)]{2pt}$\cite{Ghahremani-TVCG}\\ 
 &
  Jaccard: 
 $\tikzcircle[black,fill=electricpurple]{2pt}$\cite{Cheng-TVCG19},
 $\tikzcircle[black,fill=azure(colorwheel)]{2pt}$\cite{Kashir-JOV21}
  classification accuracy: 
 $\tikzcircle[black,fill=green(pigment)]{2pt}$\cite{Shi-TIST19}\\
  &
 classification error: 
 $\tikzcircle[black,fill=green(pigment)]{2pt}$\cite{Shi-TIST19}
   F$_{\beta}$: 
 $\tikzcircle[black,fill=azure(colorwheel)]{2pt}$\cite{Ghahremani-TVCG}
  FN: 
 $\tikzcircle[black,fill=azure(colorwheel)]{2pt}$\cite{Berenjkoub-VISSP20}
 FP: 
 $\tikzcircle[black,fill=azure(colorwheel)]{2pt}$\cite{Berenjkoub-VISSP20}\\
 &
  FPR: 
 $\tikzcircle[black,fill=azure(colorwheel)]{2pt}$\cite{Wang-VIS20}
 HR: 
 $\tikzcircle[black,fill=green-yellow]{2pt}$\cite{Hong-PVIS18}
 ROR: 
 $\tikzcircle[black,fill=electricpurple]{2pt}$\cite{Chu-TOG17}
  TN: 
 $\tikzcircle[black,fill=azure(colorwheel)]{2pt}$\cite{Berenjkoub-VISSP20}
 TP: 
 $\tikzcircle[black,fill=azure(colorwheel)]{2pt}$\cite{Berenjkoub-VISSP20}\\
 \hline
&
 kinetic energy: 
 $\tikzcircle[black,fill=richcarmine]{2pt}$\cite{Chu-TOG21}
 power spectra: 
 $\tikzcircle[black,fill=cherryblossompink]{2pt}$\cite{Wurster-arXiv21}\\
  physics &
 RMSE of vorticity: 
 $\tikzcircle[black,fill=cherryblossompink]{2pt}$\cite{Guo-PVIS20}
 RMSE of wall shear stress: 
 $\tikzcircle[black,fill=cherryblossompink]{2pt}$\cite{Guo-PVIS20} \\
 &
 TPD: 
 $\tikzcircle[black,fill=green-yellow]{2pt}$\cite{Kim-CGF19-2}
 vorticity ratio: 
 $\tikzcircle[black,fill=richcarmine]{2pt}$\cite{Chu-TOG21}\\  \hline
 human  & 
  MOS: 
 $\tikzcircle[black,fill=brinkpink]{2pt}$\cite{Han-VI22},
 $\tikzcircle[black,fill=cherryblossompink]{2pt}$\cite{Han-TVCG}
 AER: 
 $\tikzcircle[black,fill=green(pigment)]{2pt}$\cite{Yang-JOV19}\\
  &
 average top-3 match distance: 
 $\tikzcircle[black,fill=green(pigment)]{2pt}$\cite{Yang-JOV19}
 \# hits: 
 $\tikzcircle[black,fill=green(pigment)]{2pt}$\cite{Yang-JOV19} \\ \hline
\end{tabular}
}
\end{table}

\vspace{-0.1in}
\subsection{Evaluation Metrics} 
\vspace{-0.025in}

All papers we survey include qualitative results that show the visualizations generated from their solutions. 
In most cases, they also compare their works' results with other methods (including DL- and non-DL-based). 
Many of them also report the timing (including training and inference) performance of their neural networks. 
Besides qualitative results, many papers utilize quantitative metrics in their evaluations (we only find six exceptions~\cite{Xie-TOG18,Werhahn-CGIT19,He-VI20,Raji-EGPGV17,Liu-arXiv19,Han-CGF22}). 
In the following, we discuss quantitative metrics these surveyed papers employ in the evaluation. 

We categorize the evaluation metrics into six levels: {\em data}, {\em image}, {\em feature}, {\em probability}, {\em physics}, and {\em human}. 
Data-level metrics quantify the errors produced from synthesized or reconstructed data (e.g., raw scalars or vectors) compared with the ground-truth data. 
Image-level metrics compute the differences between visualization images (e.g., volume rendering, isosurface rendering, streamline visualization, pathline visualization) produced from synthesized and ground-truth data or produced from neural networks and traditional rendering processes.  
Feature-level metrics evaluate the gaps between visual representations (e.g., streamlines, pathlines, isosurfaces, stream surfaces) produced from inferred and original data.
Probability-level metrics compare the differences between predicted probabilities (e.g., boundary maps, segmentation maps) and ground-truth ones.
Physics-level metrics calculate the deviations of physics-related quantities (e.g., power spectra, kinetic energy) derived from synthesized and ground-truth data.
Finally, human-level metrics ask human subjects to give ratings or scores to the results (typically visualization results) produced from synthesized data with ground-truth references or compare results (e.g., viewpoints) suggested by neural networks with those selected by humans. 

As shown in Table~\ref{tab:metrics}, data-level metrics are most popular, followed by feature-, probability-, and image-level metrics. Physics- and human-level metrics are the least used ones. 
Across the research tasks, data generation tasks [$\tikzcircle[black,fill=cherryblossompink]{2pt}$ $\tikzcircle[black,fill=brinkpink]{2pt}$ $\tikzcircle[black,fill=richcarmine]{2pt}$ $\tikzcircle[black,fill=carmine]{2pt}$] use metrics across all but probability-level metrics. 
However, visualization generation tasks [$\tikzcircle[black,fill=gold(web)(golden)]{2pt}$] exclusively utilize image-level metrics. 
Data-relevant prediction tasks [$\tikzcircle[black,fill=green-yellow]{2pt}$] use data-, feature-, probability, and physics-level metrics, while visualization-relevant prediction tasks [$\tikzcircle[black,fill=green(pigment)]{2pt}$] utilize feature-, probability-, and human-level metrics. 
Most object detection and segmentation tasks [$\tikzcircle[black,fill=azure(colorwheel)]{2pt}$] employ probability-level metrics (with a few exceptions using image- and feature-level metrics). 
Finally, feature learning and extraction tasks [$\tikzcircle[black,fill=electricpurple]{2pt}$] utilize data-, feature-, and probability-level metrics 
but no image- and human-level metrics. 

PSNR, SSIM, IS/MSE, and F-score are the most widely used ones in the data-, image-, feature-, and probability-level metrics, respectively. In addition, several metrics (i.e., PSNR, SSIM, MSE, RMSE, MAE, EMD, F-score, IOU) are utilized across different categories. A closer look shows that many papers employ more than one evaluation metric, and in this case, several of them (e.g.,~\cite{Han-VIS19,Guo-PVIS20,Chu-TOG21,An-CGA21}) utilize metrics across categories for a comprehensive evaluation. 

Several data- and feature-level metrics are primarily used for vector fields, flow lines, or critical points. 
AAD and CD$^2$ 
are data-level metrics for comparing individual vectors' angles and magnitudes. 
At the feature level, ALP and MCPD 
compare integral flow lines' endpoints or calculate the distances among sample points along flow lines. 
ACPPE considers critical point position deviations. 
In addition, three metrics (i.e., LSiM, VGG metric, LPIPS) are learned metrics based on neural networks.  

Most data-level metrics evaluate the data error at individual voxel or pixel (i.e., 2D data slice) levels. 
Other than that, CR is often used to evaluate data compression or reduction performance. 
LSiM~\cite{Kohl-ICML20} was utilized to evaluate the accuracy of static and temporal restoration in fluid simulation~\cite{Chu-TOG21}. 
VGG metric~\cite{Simonyan-ICLR15} was compared against self-supervised learning of spatiotemporal similarity~\cite{Tkachev-TVCG}.

For image-level metrics, 
PSNR, MAE, and RMSE 
examine the differences of two images at the pixel level. 
SSIM compares two images according to the patch-level means, variances, and covariances. 
As a network-based similarity metric, LPIPS~\cite{Zhang-CVPR18} computes a weighted average of the activations at hidden layers to predict relative image similarities correlating well with perceptual judgments. 
EMD~\cite{Rubner-IJCV00} and FID~\cite{Heusel-NIPS17} quantify the distances between two images at the histogram and distribution levels. 

Two feature-level metrics were used to evaluate surface similarities: IS~\cite{Bruckner-CGF10} (which employs MI to identify the similarity between isosurfaces) and surface-based HD~\cite{Wiewel-CGF19} (which uses HD to compute the error between the predicted and reference liquid surfaces represented in signed distance functions).  

Several probability-level metrics are related to the confusion matrix in a supervised learning setting (e.g., classification or segmentation task). The basic terms of the confusion matrix are FN, FP, TN, and TP, and their derivations include accuracy, F$_{\beta}$, FPR, F-score, IOU, Jaccard, PPV, and TPR. 

Finally, we notice that physics- and human-level metrics are far less employed in the evaluation. This is due to the lack of physics-informed DL works in SciVis and the missing of a rich set of human-level quantitative metrics.

\vspace{-0.1in}
\section{Research Opportunities}
\vspace{-0.025in}

DL4SciVis is a fast-growing area in SciVis, which may play a pivotal role in the future of SciVis. 
Reflecting on several dimensions (i.e., domain setting, research task, learning type, network architecture) used to classify the surveyed papers (refer to Tables~\ref{tab:papers},~\ref{tab:learning}, and~\ref{tab:network}), we can identify gaps that suggest possible future research directions. 
This section examines the remaining gaps based on the surveyed papers and points out several research opportunities for DL4SciVis. 

{\bf From single field to multiple fields.}
Table~\ref{tab:papers} shows that more than half of the surveyed papers are for scalar field data, which is about twice as many as those for vector field data. Although many papers deal with time-varying scalar fields (e.g.,~\cite{Han-VIS19,Han-TVCG,Han-VIS21,Tkachev-TVCG21,Tkachev-TVCG}) or unsteady vector fields (e.g.,~\cite{An-CGA21,Jakob-VIS20,Gu-CGA21,Hong-PVIS18,Kim-CGF19-2}), only a few works handle multivariate data~\cite{Han-VIS20,Porter-VISSP19} or ensemble data~\cite{He-VIS19,He-VI20}. 
We expect the future growth of DL4SciVis by considering multi-field (scalar, vector, tensor) and multi-run (ensemble) data and the interplay among them. For example, Chu et al.\ \cite{Chu-TOG21} considered the data translation problem (from density scalar field to velocity vector field). 
In CV, DL-based image-to-image translation (e.g., Pix2Pix~\cite{Isola-CVPR17}, CycleGAN~\cite{Zhu-ICCV17}) and image colorization~\cite{Zhang-ECCV16} works provide us good examples to design suitable solutions for SciVis data. 

{\bf From data generation to visualization generation.}
In Table~\ref{tab:papers}, we can see a starking contrast between the numbers of papers on data generation and visualization generation (even including visualization-relevant prediction works). Data generation works have grown significantly in the subcategories of super-resolution generation and compression and reconstruction. Visualization generation, however, presents more challenges as we need to consider different parameters involved in the rendering process, including transfer function, viewpoint, lighting, etc. Therefore, the lagging of visualization generation behind data generation is reasonable. Nevertheless, the new surge of differentiable rendering~\cite{Kato-arXiv20} or neural rendering~\cite{Tewari-CGF20} in CG could be poised to become a new area in SciVis. 

{\bf From images and volumes to graphs.} 
We can observe from Tables~\ref{tab:papers-IO-1} to~\ref{tab:papers-IO-5} that most papers focus on processing SciVis data in their original forms (e.g., images and volumes) while derived forms (e.g., graphs) are seldom operated. So far, only four works use GNN~\cite{He-JOV22-1,He-JOV22-2,Han-CGF22,Shi-PVIS22}, where GCNs are employed to learn scalar value association, super-voxel features, surface node features, and feature map upsampling, respectively. GNN is not limited to GCN~\cite{Zhang-CSN19}, and it also includes GAE, GRN, and STGNN~\cite{Wu-TNNLS21}. SciVis data have rich graph-like representations or relationships~\cite{Wang-CGF17}, such as surfaces (e.g., isosurfaces, stream surfaces) and relationships (e.g., correlation, transition, topology). Advances in GNN techniques from CG (e.g., geometric DL~\cite{Bronstein-SPM17} on non-Euclidean domains such as graphs and manifolds) and knowledge discovery and data mining (KDD)~\cite{Zhou-AIO20,Wu-TNNLS21,Zhang-TKDE22} will provide ample opportunities for us to develop GNN-based solutions for solving SciVis problems.  

{\bf From supervised learning to self-supervised learning.}
In Table~\ref{tab:learning}, we can see that the majority of the surveyed papers fall into the category of supervised learning, where a large amount of annotated data is needed for training. However, this requirement is not easy to meet when the annotation is time-consuming and difficult to collect. In CV, self-supervised learning was proposed to address this issue. Self-supervised learning designs a pretext task to learn hidden representations with a large amount of unlabeled data. It then fine-tunes a downstream task (e.g., classification and segmentation) with few annotated data. Although there are several self-supervised learning works (i.e.,~\cite{Chu-TOG17,Raji-EGPGV17,Tkachev-TVCG21}) in SciVis, they all utilize the Siamese network~\cite{Bromley-NIPS93} for pairwise contrastive learning. Furthermore, these works focus on feature extraction without extending the framework to fine-tune the downstream tasks. In addition, the relationship between pretext and downstream tasks is still unclear. We expect more works in this category, especially leveraging new frameworks (such as CMC~\cite{Tian-ECCV20}, SimCLR~\cite{Chen-ICML20}, MoCo~\cite{He-CVPR20}, and BYOL~\cite{Grill-NIPS20}) and novel pretext tasks. Besides contrastive learning, the unexplored generative learning~\cite{Liu-TKDE} presents a unique opportunity. 

{\bf From CNN and RNN to MLP.}
We can see from Table~\ref{tab:network} that CNNs and RNNs have been well utilized in DL4SciVis research. They have also been extensively used in generative tasks, such as super-resolution generation and visualization generation. To our surprise, pure MLP was seldom employed in these tasks, and the only MLP work (i.e.,~\cite{Lu-CGF21}) is for data compression. Recent investigations in CV have demonstrated that a pure MLP model can outperform CNN- and RNN-based architectures across a diverse set of tasks, such as image classification~\cite{Dosovitskiy-ICLR21,Liu-ICCV21}, image segmentation~\cite{Zheng-CVPR21}, object detection~\cite{Carion-ECCV20}, image reconstruction~\cite{Sitzmann-NIPS20}, and view synthesis~\cite{Chan-CVPR21}. We expect more DL4SciVis works utilizing MLPs and newer mechanisms or architectures, such as transformer~\cite{Vaswani-NIPS17} and implicit neural representation~\cite{Mildenhall-ECCV20}. 

{\bf From distributed learning to disentangled learning and style transfer.}
We observe that most unsupervised learning works in SciVis are distributed learning, e.g., an encoder+decoder network architecture is designed to extract unified features through reconstruction. However, the possibility of disentangled learning (refer to Figure~\ref{fig:learning}) and style transfer is still unexplored. In CV, Gatys et al.\ \cite{Gatys-arXiv15} pioneered the use of CNNs to extract feature responses of a photo as the {\em content} and the feature statistics of a piece of artwork as the {\em style} for neural style transfer (i.e., rendering a content image in different styles). Huang et al.\ \cite{Huang-ECCV18} built two encoders to learn the content (e.g., shape) and style (e.g., texture and pose) of an image, respectively. After that, fixing the content feature while switching different style features can render arbitrary images with the same content in different styles. Many styles can be considered, including semantic, instance, doodle, stereoscopic, portrait, video, character, photorealistic, attribute, fashion, and audio styles~\cite{Jing-TVCG20}. Although it is not as intuitive as images or videos to define content and style for SciVis data, it is not impossible. For example, given an ensemble fluid flow simulation, the content could be invariant information (e.g., vortices). The style could be the pattern (e.g., the number, size, and location of vortices) extracted by simulation output under varying Reynolds numbers, which indicate how turbulent the flow is. If such a definition is meaningful and the transferred results are interpretable, disentangled learning for data generation can play a role in DL4SciVis. 

{\bf From heavyweight to lightweight.} 
Currently, the DL models in DL4SciVis are built with tens or hundreds of layers to guarantee quality. However, this could result in a large model size and inefficient inference. In the ML community, researchers have already studied different techniques (e.g., weight quantization~\cite{Han-ICLR16,Han-HCS16,Lin-ICLR19} and knowledge distillation~\cite{Hinton-arXiv15,Li-CVPR20,Fang-ICCV21,Zhu-ICCV21}) to build a lightweight model from a heavyweight one. For instance, Han et al.\ \cite{Han-ICLR16} pruned the network, quantized parameters and compressed them using Huffman coding. Li et al.\ \cite{Li-CVPR20} applied neural architecture search to find efficient architectures through combining the knowledge of multiple intermediate features extracted from the heavyweight model. Thomas et al.\ \cite{Thomas-TOG20} presented QW-Net for image reconstruction, where about 95\% of the computations can be implemented with 4-bit integers. We believe there is an opportunity to incorporate these techniques into DL models to improve training efficiency for large-scale scientific data analysis and visualization.

{\bf From centralized learning to federated learning.}
The success of DL models heavily relies on a large amount of data. However, due to practical issues such as confidentiality and privacy, SciVis data are often not publicly available. This prevents the DL models from gaining a strong learning capability from different sources. Instead of requiring data sets, researchers can release the trained models. Federated learning~\cite{Kairouz-FTML21} can produce a shared model by collaborating with local models trained on different data sets. The shared model does not need to access the trained data sets. Techniques in federated learning include weight averaging~\cite{Mcmahan-ICAIS17}, momentum update~\cite{Hsu-arXiv19}, Bayesian non-parametric match~\cite{Wang-ICLR20}, and model contrast~\cite{Li-CVPR21-2}. Although these algorithms were designed for classification tasks, we expect researchers to explore this direction in SciVis-related tasks by designing new approaches when multiple local models are available.

{\bf From data-driven DL to physics-informed DL.} 
Existing works in DL4SciVis are primarily data-driven, require a significant amount of data for network training. 
These works are often purely data-driven, seldom leveraging the underlying physics for physics-informed DL. 
Outside the SciVis community, researchers in CFD and fluid simulation have extensively investigated physics-informed DL~\cite{Karniadakis-NRP21,Thuerey-arXiv21}. 
For instance, Raissi et al.\ \cite{Raissi-JCP19} introduced a physics-informed neural network for solving supervised learning tasks involving nonlinear partial differential equations. 
Considering the data and physics scenarios, we have one extreme of big data with no physics and the other extreme of small data with lots of physics. 
Physics-informed DL applies to the middle regime with some data and some physics. 
Using differentiable physics and neural networks, physics-informed DL can integrate data and the governing physical laws to produce predictions conforming to the underlying physics, even for models with partially missing physics. 
As listed in Table~\ref{tab:papers}, several surveyed papers in the fluid simulation domain (i.e.,~\cite{Xie-TOG18,Werhahn-CGIT19,Kim-CGF19-1,Chu-TOG21,Wiewel-CGF19,Wiewel-CGF20}) have considered physical information or quantities in their works, sharing some flavor of physics-informed DL. Although there are practical gaps to be bridged, SciVis researchers may find it rewarding to work closely with physics and simulation researchers to jointly advance physics-informed DL.

\vspace{-0.1in}
\section{Open Challenges}
\vspace{-0.025in}

In a recent article published in {\em Communications of the ACM}, Bengio et al.\ \cite{Bengio-CACM21} outlined some of the future challenges facing DL for AI, including training with little or no supervision, robustness to test samples out of training data distribution, and DL for tasks requiring a deliberate sequence of steps. 
Compared to the advances of DL in CV and CG, DL4SciVis is still in its early stage of development. 
Despite significant progress over the past few years, many breakthroughs still need to be made in DL4SciVis research. 
In this section, we identify and discuss several open yet pressing challenges facing us.

{\bf Model generalization.}
A universal neural network model trained on various images could effectively perform intended tasks on multiple image categories. However, this is not the case for SciVis data due to the lack of sufficiently large and diverse enough training data and the added training cost (2D images vs. 3D volumes). Current DL4SciVis works allow training a model on specific variables or ensemble runs and later applying the model to a different variable sequence~\cite{Han-VIS20} or ensemble run~\cite{He-VIS19} of the same simulation. Beyond that, the performance often downgrades~\cite{Han-TVCG20}. SciVis researchers have not seriously investigated the issue of generalizing the neural network model to adapt appropriately to new, previously unseen data. The current practice of ``one training for one dataset'' ought to be changed. Training a generalized model to work well from one dataset to different datasets poses an open challenge. Commonly used in CV, data augmentation techniques~\cite{Shorten-JBD19} can help improve model generalization. Their goal is to increase the amount of training data by adding modified copies of existing data (e.g., via cropping, transformation, noise injection, and random erasing) or newly created synthetic data from existing data using generative solutions~\cite{Li-CVPR21-1,Zhang-CVPR21}. Besides training with large and diverse data, it is also possible to improve model generalization by introducing pre-training stages~\cite{Erhan-JMLR10} 
or building connections between early and latter layers (e.g., residual connection~\cite{He-CVPR16} and dense connection~\cite{Huang-CVPR17}). 

{\bf Benchmark dataset.} 
CV and CG researchers have produced many benchmark datasets for reproducible research. The hugely successful benchmark databases such as ImageNet~\cite{Deng-CVPR09} demonstrate their value of benefiting the community and advancing the field. In SciVis, such an effort is scarce. SciVis researchers are often short of experimental data. Unlike CV and CG, where images, videos, and geometric models are readily available for use, SciVis researchers only obtain datasets from domain scientists they collaborate with or from publicly available sources (e.g., IEEE SciVis Contest), which are rather limited. Recently, Eckert et al.\ \cite{Eckert-TOG19} produced a large-scale volumetric dataset of scalar transport flows for computer animation and ML. Jakob et al.\ \cite{Jakob-VIS20} released a large numerical 2D fluid dataset, including laminar and turbulent flows, for ML. These encouraging initiatives will inspire more significant community efforts to propel DL4SciVis and SciVis research in general. 

{\bf Few-shot learning.}
DL tasks are data-hungry. For example, generating super-resolution for time-varying volumetric data needs 40\% of samples from the datasets for training~\cite{Han-VIS19,Han-TVCG} and training a generative model for volume rendering requires 200,000 image samples~\cite{Berger-TVCG19}. In CV, Li et al.\ \cite{Li-TPAMI06} introduced one-shot learning, a Bayesian approach for learning object categories, where much information about a new object category can be learned from a single or just a few training examples. Leverages prior knowledge, few-shot learning can generalize to new tasks involving only a small number of samples with supervised information~\cite{Wang-CS20}. This direction is especially appealing to DL4SciVis research due to the limited data and time-consuming training. However, it remains challenging to determine what prior knowledge can be obtained and how such knowledge should be utilized in few-shot learning.  

{\bf Multi-task learning.}
All surveyed papers we survey only tackle a single task. Multi-task learning~\cite{Caruana-ML97} aims to learn multiple related tasks {\em simultaneously} by sharing the knowledge obtained from different tasks to improve the generalization performance of all these tasks. Zhang and Yang~\cite{Zhang-TKDE21} classified different multi-task learning algorithms into the following categories: feature learning, low-rank, task clustering, task relation learning, and decomposition. Existing DL works on multi-tasking learning are often based on sharing hidden layers of the neural network, which is vulnerable to noisy and outlier tasks. For DL4SciVis, open questions include defining and group tasks, exploiting unrelated tasks, and applying multi-task learning in non-supervised learning scenarios. 

{\bf Interpretable DL.}
All the surveyed works treat the DL models as black boxes. This makes it difficult to interpret or modify the model results when the predictions are inaccurate or do not meet particular constraints (e.g., the cell size in medical images and physical properties in simulation data). Interpretable DL~\cite{Zhang-FITEE18} aims to study the role of each neuron in a DL model and understand the decision process. Recent works~\cite{Bau-ICLR19,Bau-TOG19,Bau-ECCV20,Bau-PNAS20} investigated the importance of each neuron in image classification and generation tasks (e.g., identifying the neurons that can control the generation or classification of church). With this interpretation, researchers can manipulate the model behavior to generate desired results. For example, a GAN model can produce images without a sofa by eliminating the sofa neurons. For DL4SciVis, open questions include 
discovering the neurons with different roles, controlling specific neurons, and rewriting the neurons to produce customized results. 

{\bf Automated ML.} 
Most DL4SciVis works entail great efforts on researchers. They need to be involved in every stage of the process, including problem definition, data collection, feature engineering, model selection, algorithm selection, evaluation, and deployment. Automated ML~\cite{Yao-arXiv18} can relieve us from intermediate steps (i.e., feature engineering, model selection, algorithm selection, and evaluation), minimizing human participation and improving efficiency. Of particular interest for SciVis researchers is neural architecture search~\cite{Zoph-ICLR17,Liu-ECCV18}. Designing network architectures has been the primary task for achieving good learning performance, often demanding a time-consuming and painstaking process. For example, a typical CNN design space includes many choices, such as the number of filters, filter width and height, stride width and height, and skip connections. The iterative nature of the architecture generation process makes reinforcement learning~\cite{Sutton-BOOK98} a suitable choice for neural architecture search. Open questions for DL4SciVis researchers include determining the search space, the corresponding feedback, and the number of configurations for evaluation. 

\vspace{-0.1in}
\section{Conclusions}
\vspace{-0.025in}

We have presented a state-of-the-art survey on DL4SciVis. The survey covers 59 papers published since 2017 along six dimensions, provides an in-depth discussion on their similarities and differences, identifies trends and gaps, and outlines research opportunities and open challenges. Despite the fantastic advances of DL4SciVis, we acknowledge that researchers need to address many practical issues to make the presented solutions robust to outliers, generalizable across datasets, and applicable in real-world settings. As DL4SciVis has grown out of its infancy stage, we anticipate future research will answer these challenges. 

DL4SciVis is only a branch of AI4VIS. With AI4VIS and VIS4AI, the entire area of AI+VIS has quickly become the most vibrant research focus in VIS. The astonishing advancement of ML and DL, the interplay between AI4VIS and VIS4AI, and the interconnection across SciVis, InfoVis, and VA provide a myriad of thoughts and ideas for sustainable growth of AI+VIS research for years to come. We hope this survey can serve as a good source of reference for SciVis researchers and shed light on future directions.

\vspace{-0.1in}
\section*{Acknowledgements}
\vspace{-0.025in}

This work was supported in part by the U.S.\ National Science Foundation through grants IIS-1455886, CNS-1629914, DUE-1833129, IIS-1955395, IIS-2101696, and OAC-2104158. The authors would like to thank the anonymous reviewers for their insightful comments.

\vspace{-0.1in}
\bibliographystyle{abbrv}
\bibliography{template-collection,template-deep-learning,template-others,template-jun}

\begin{IEEEbiography}[{\includegraphics[width=1in,height=1.25in,clip,keepaspectratio]{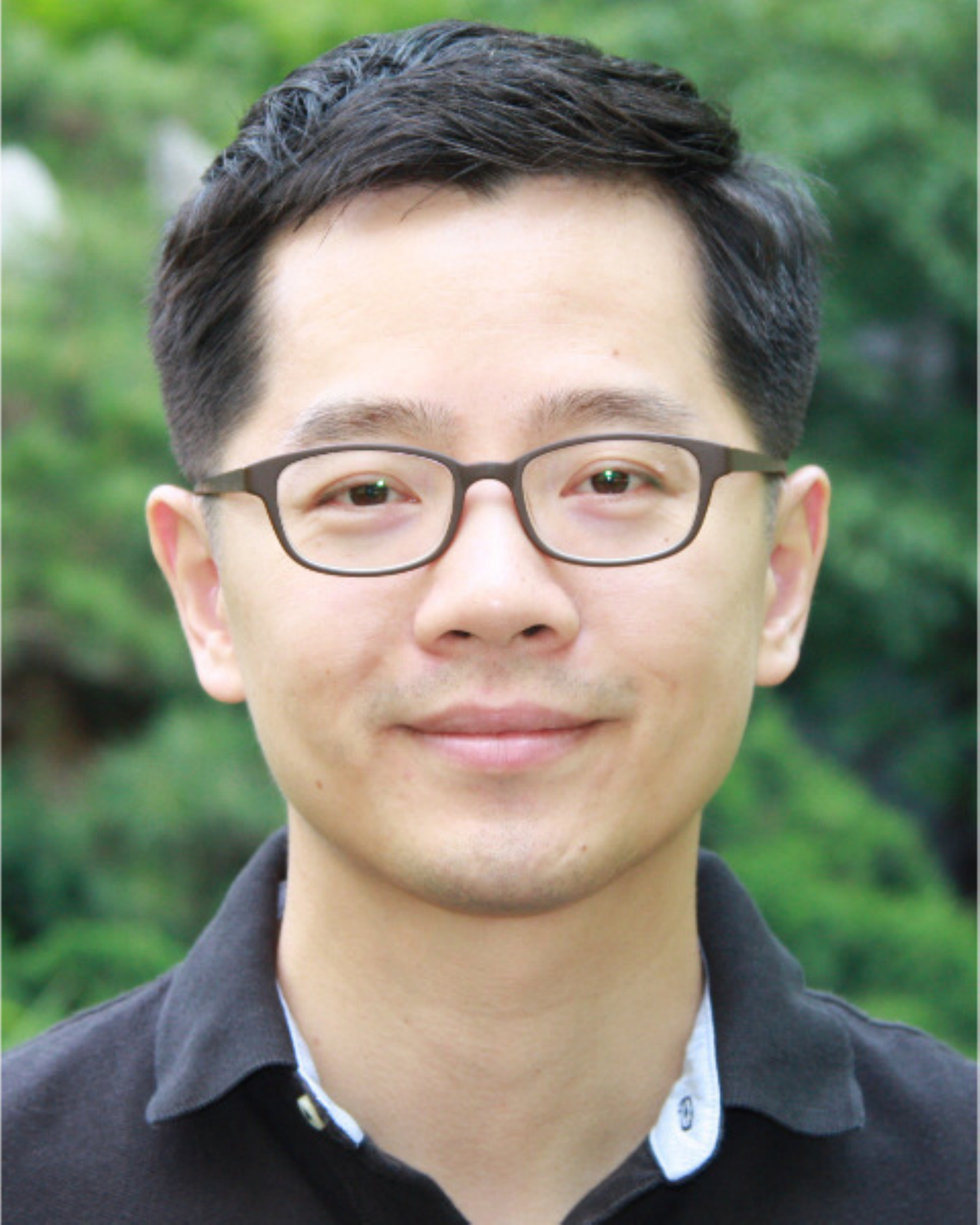}}]{Chaoli Wang} is an associate professor of computer science and engineering at the University of Notre Dame. He received a Ph.D. degree in computer and information science from The Ohio State University in 2006. Dr.\ Wang's primary research interest is data visualization, particularly on the topics of time-varying multivariate data visualization, flow visualization, information-theoretic algorithms, graph-based techniques, and deep learning solutions for big data analytics. He is an associate editor of IEEE Transactions on Visualization and Computer Graphics. 
\end{IEEEbiography}

\begin{IEEEbiography}[{\includegraphics[width=1in,height=1.25in,clip,keepaspectratio]{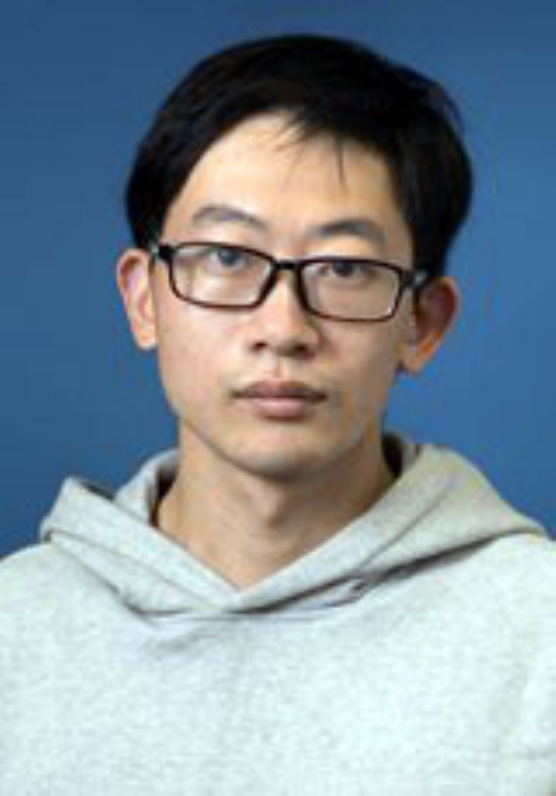}}]{Jun Han} is a Ph.D.\ candidate at the University of Notre Dame. He received a BS degree in software engineering and an MS degree in computer software and theory in 2014 and 2017. Both degrees are from Xidian University. His Ph.D. research focuses on applying deep learning techniques to solve data visualization problems.
\end{IEEEbiography}

\end{document}